\documentclass[10pt,twocolumn]{article} 
\usepackage{booktabs} 
\usepackage{amsmath,amsfonts,amssymb,amsthm}
\usepackage[linesnumbered,ruled,noend]{algorithm2e}
\usepackage{graphicx}
\usepackage{textcomp}
\usepackage{xcolor}
\usepackage{adjustbox}
\usepackage{multirow}
\newtheorem{definition}{Definition}[section]
\usepackage{subcaption}
\newtheorem{lemma}{Lemma}
\newtheorem{theorem}{Theorem}
\usepackage{comment}
\date{}

\begin{document}

\title{The GraphTempo Framework for Exploring the Evolution of a Graph through Pattern Aggregation}

\author{
  Evangelia Tsoukanara\\
  Department of Applied Informatics,\\
  University of Macedonia, Greece\\
  etsoukanara@uom.edu.gr
  \and
  Georgia Koloniari\\
  Department of Applied Informatics,\\
  University of Macedonia, Greece\\
  gkoloniari@uom.edu.gr
  \and
  Evaggelia Pitoura\\
  Department of Computer Science \& \\
  Engineering, University of Ioannina, Greece\\
  pitoura@cse.uoi.gr
  \and
  Peter Triantafillou\\
  Department of Computer Science,\\
  University of Warwick, Coventry, UK\\
  pitoura@cse.uoi.gr
}

\maketitle
\thispagestyle{empty}

\begin{abstract}
When the focus is on the relationships or interactions between entities, graphs offer an intuitive model for many real-world data. Such graphs are usually large and change over time, thus, requiring models and strategies that explore their evolution. We study the evolution of aggregated graphs and introduce the GraphTempo model that allows temporal and attribute aggregation not only on node level by grouping individual nodes, but on a pattern level as well, where subgraphs are grouped together. Furthermore, We propose an efficient strategy for exploring the evolution of the graph based on identifying time intervals of significant growth, shrinkage or stability. Finally, we evaluate the efficiency and effectiveness of the proposed approach using three real graphs.
\end{abstract}

\section{Introduction}
Graphs provide an intuitive model for representing relationships and interactions between entities. Their use expands a variety of domains ranging from social and contact networks to road and biological networks. Most such graphs are usually large and dynamic, thus requiring efficient models to capture their evolution and identify trends or events of interest. 

Instead of studying the evolution of individual graph nodes, we consider aggregate graphs that offer a summarized view of the network and allow us to study evolution at a global level. There has been a lot of work on aggregation of static graphs e.g., \cite{Zhao11, Chen09}, and more recently on dynamic graphs, where problems such as alternative temporal graph models \cite{Ghrab13}, \cite{Moffitt17}, \cite{Aghasadeghi20}, temporal paths evaluation \cite{Debrouvier21}, graph analytics \cite{Guminska18} and visualization \cite{Rost21} have been addressed. In this paper, we focus on aggregate graph evolution and propose novel strategies for detecting periods of significant stability, growth or shrinkage. 

We consider temporal attributed graphs where nodes have properties 
whose values may or not change with time, and introduce  the \textit{GraphTempo} model for aggregating such graphs and exploring their evolution.  Our model supports three types of aggregation: \textit{temporal} aggregation that enables us to study a graph at different levels of temporal resolution, \textit{attribute} aggregation that groups nodes that share common attribute values and, \textit{pattern} aggregation that extends attribute aggregation by grouping together subgraphs. 

To motivate our work, consider a contact network modeling the interactions of students and teachers to study the spread of an infectious disease \cite{Gemmetto14}. As students tend to interact with others of their own grade and mostly within their class, studying the aggregated graph based on time and class attribute enables us to identify stable relationships between students of the same class that can be seen as isolation bubbles that hinder disease spread, while periods showing growth between interaction of students of different classes identifies time periods, such as breaks that pose a higher risk in disease spread. Furthermore, extending our study to larger groups of students, for instance by considering three students that form a pattern of a closed triangle with their interactions and aggregating the graph by grouping triangles of students of the same class, can offer us further intuitions on the size of the isolation bubbles. 

To support temporal aggregation we define set based temporal operators using either tight \textit{intersection} semantics  or more relaxed \textit{union} semantics. We discern between distinct and non-distinct aggregation and provide algorithms for all types of aggregation. We also propose optimizations that reuse evaluated aggregate graphs to compute other aggregate graphs more efficiently without having to access the original temporal graphs.

With regards to evolution, we define three types of events pertaining to the \textit{growth}, \textit{stability} and \textit{shrinkage} of the graph. We define the \textit{evolution graph} that captures the three types of evolution events by overlaying three different graphs defined using appropriate temporal operators for each event type. To explore graph evolution, we define the novel problem of finding pairs of intervals $(\mathcal{T}_1, \mathcal{T}_2)$ such that at least $k$  events of interest have occurred between $\mathcal{T}_1$ and $\mathcal{T}_2$. Since the number of candidate interval pairs can grow exponentially, we propose an efficient exploration strategy that considers a fixed reference point and utilizes intersection and union semantics to extend intervals while exploiting monotonicity to further prune the exploration space.
Our experimental evaluation utilizes three diverse real world datasets to showcase both the efficiency and the effectiveness of the proposed approach. 

The rest of the paper is structured as follows. In Section 2, we define the three types of graph aggregation, and the evolution graph, and in Section 3, our problem of exploring graph evolution and the exploration strategies. In Section 4,  we present our algorithms, and in Section 5 our experimental results. Section 6 summarizes related work, while Section 7 offers conclusions.

\section{The GraphTempo Model}
To define a temporal graph model, we first consider a linearly ordered discrete time domain consisting of time points $t$, where the linear ordering signifies that if $t_i<t_{i+1}$ then $t_i$ happens before $t_{i+1}$. Given $t_i,t_j, t_i<t_j$, a time interval $\mathcal{T}$ is defined as $\mathcal{T}=[t_i,t_j]$, and includes all times points $t$ from $t_i \leq t \leq t_j$. The length of the interval, $length(\mathcal{T})$ is defined as the number of time points $t \in \mathcal{T}$. We assume an interval-based definition, where an attributed temporal graph in a time interval $\mathcal{T}$ is a graph whose elements, i.e., nodes, edges and attributes are annotated with the set of intervals in  $\mathcal{T}$ during which they were valid.
 \begin{definition} [Temporal Attributed Graph]
	A temporal attributed graph in $\mathcal{T}$ is a graph $G(V, E, \tau u, \tau e, A)$ where $V$ is the set of nodes and  $E$ is the set of edges $(u, v)$ with $u, v$ $\in$ $V$.
	Each node in $V$ is associated with a timestamp  $\tau u: V \rightarrow \mathcal{T}$ where $\tau u(u)$ is the set of intervals during which $u$ exists.
	Similarly each edge $e$ in $E$ is associated with a timestamp  $\tau e: E \rightarrow \mathcal{T}$ where $\tau e(e)$ is the set of intervals during which $e$ exists. $A$ is a set of arrays,  one for each of the set of $n$ attributes associated with vertices of $V$. That is, for each $u \in V$ and $t \in \tau u(u)$, there is an array, $A(u, t) = \{A^1(u, t), A^2(u, t) \dots A^k(u, t)\}$, where $A^i(u, t)$ denotes the value of $u$ at time $t$ $\in \tau u(u)$  on the $i$-th attribute. 
\end{definition}
An attribute $A^i$ is called \textit{static}, if its value does not change with time, i.e., $A^i(u, t) = A^i(u, t')$,  $\forall u \in V$ and $\forall t,t' \in \tau(u) \subseteq T$, and \textit{time-varying} otherwise.

\begin{figure}
\centering
\includegraphics[scale=0.5]{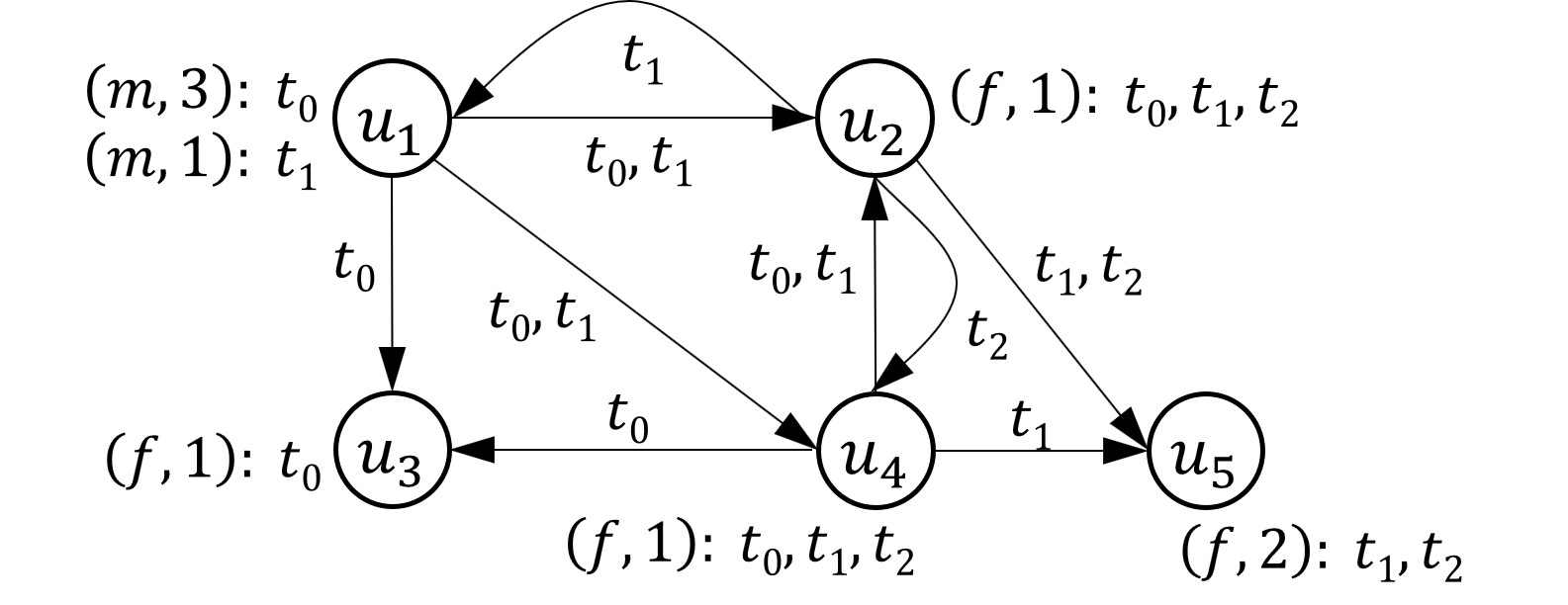}
\caption{A temporal attributed graph.}
\label{fig:1}
\end{figure}

Figure \ref{fig:1} depicts an example of a temporal co-authorship attributed graph in time interval $\mathcal{T}=[t_0,t_2]$. The nodes have two attributes, ``Gender'', a static categorical attribute with values $\{m,f\}$ and ``\#Publications'', a time-varying numerical attribute. 

\begin{figure}
\centering
\includegraphics[scale=0.5]{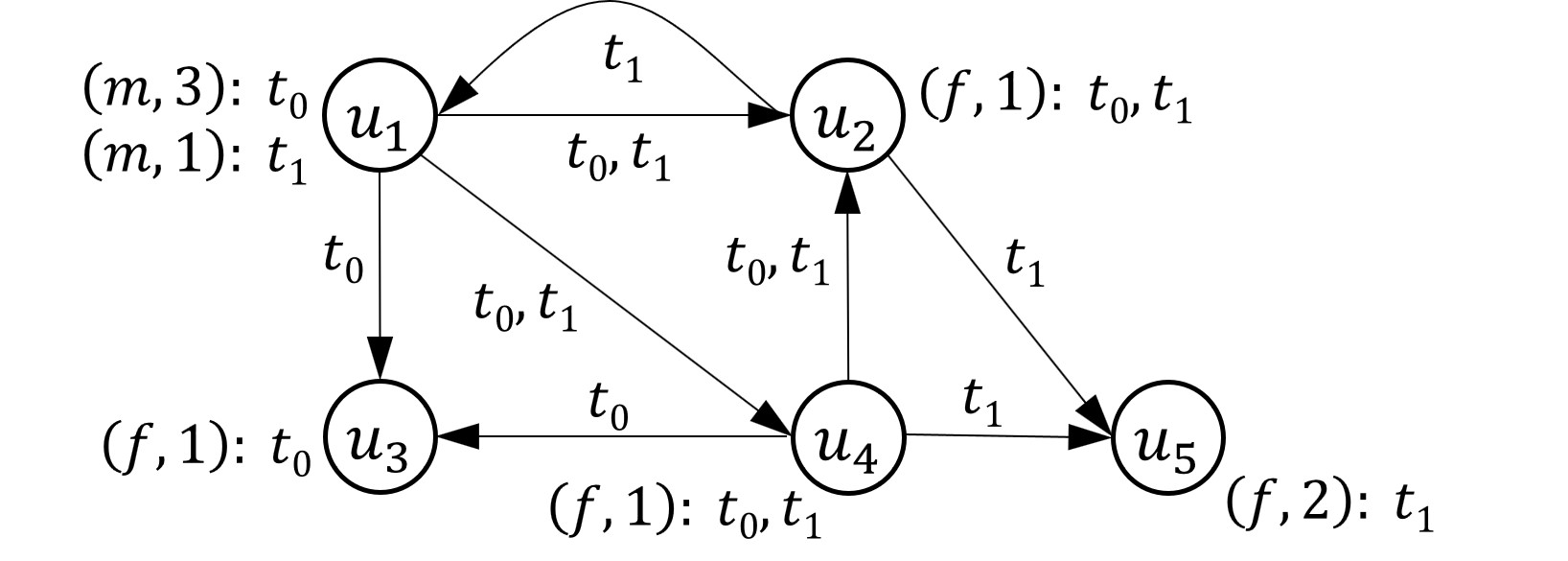}
\caption{Union graph of the graph of Fig. \ref{fig:1} in $[t_0, t_1]$.}
\label{fig:335}
\end{figure}

\subsection{Temporal Aggregation}
To define temporal attributed graphs in intervals of different length and combinations of intervals, we introduce \textit{temporal aggregation}. 

We first define an auxiliary operator, \textit{project}, that defines a subgraph of the original temporal graph in specific time intervals.
\begin{definition} [Time Project Operator]
	The projection $G[\mathcal{T}_1]$ of $G$ in $\mathcal{T}_1 \subseteq \mathcal{T}$ is a temporal attributed graph in $\mathcal{T}_1$ $G[\mathcal{T}_1](V_1, E_1,\tau u_1, \tau e_1, A_1)$, where 
	$V_1$ includes all nodes $u \in V$ for which ${\mathcal{T}_1} \subseteq \tau u(u)$,
	$E_1$ all edges $e \in E$ for which ${\mathcal{T}_1} \subseteq \tau e(e)$, and $\forall u \in V_1$, $\tau u_1(u)=\tau u(u) \cap \mathcal{T}_1$, and  $\forall e \in E_1$, $\tau e_1=\tau e(e) \cap \mathcal{T}_1$ and $A_1 \subseteq A$ includes all $A(u, t)$ for which $u \in V_1$ and $t \in \tau u_1(u)$.
\end{definition}
We provide operators that given two sets of time intervals $\mathcal{T}_1$ and $\mathcal{T}_2$, they output the graph with all elements that exist in $\mathcal{T}_1$ or $\mathcal{T}_2$ (union), in both $\mathcal{T}_1$ and $\mathcal{T}_2$ (intersection), or in $\mathcal{T}_1$ and not $\mathcal{T}_2$ (difference). In particular:
\begin{itemize}  
    \item \textbf{union ($\cup$)} defines graph $G[\mathcal{T}_1 \cup \mathcal{T}_2](V_{\cup}, E_{\cup},\tau u_{\cup},$ $\tau e_{\cup}, A_{\cup})$, where 
$V_{\cup} = \{u | (\tau u(u) \cap (\mathcal{T}_1 \cup \mathcal{T}_2) \neq \emptyset \}$, 
$E_{\cup} = \{e | (\tau u(u) \cap (\mathcal{T}_1 \cup \mathcal{T}_2) \neq \emptyset \}$, $\forall u \in V_{\cup}$,  $\tau u_{\cup}(u)=\tau u(u) \cap (\mathcal{T}_1 \cup \mathcal{T}_2)$, $\forall e \in E_{\cup}$, $\tau e_{\cup}(e)=\tau e(e) \cap (\mathcal{T}_1 \cup \mathcal{T}_2)$ and $A_{\cup}$ includes all $A(u, t)$ for which $u \in V_{\cup}$ and $t \in \tau u_{\cup}(u)$.
\item \textbf{intersection ($\cap$)} defines graph $G[\mathcal{T}_1 \cap \mathcal{T}_2](V_{\cap}, E_{\cap},$ $ \tau u_{\cap},\tau e_{\cap}, A_{\cap})$, where $V_{\cap} = \{u | (\tau u(u) \cap \mathcal{T}_1 \neq \emptyset) \textit{ and } (\tau u(u) \cap \mathcal{T}_2 \neq \emptyset)  \}$, $E_{\cap} = \{e | (\tau e(e) \cap \mathcal{T}_1 \neq \emptyset) \textit{ and } (\tau e(e) \cap \mathcal{T}_2 \neq \emptyset)  \}$, $\forall u \in V_{\cap}$,  $\tau u_{\cap}(u)=\tau u(u) \cap (\mathcal{T}_1 \cup \mathcal{T}_2)$, $\forall e \in E_{\cap}$, $\tau e_{\cap}=\tau e(e) \cap (\mathcal{T}_1 \cup \mathcal{T}_2)$ and $A_{\cap}$ includes all $A(u, t)$ for which $u \in V_{\cap}$ and $t \in \tau u_{\cap}(u)$.
 \item \textbf{difference ($-$)} between $\mathcal{T}_1$ and $\mathcal{T}_2$ that defines graph $G[\mathcal{T}_1-\mathcal{T}_2](V_{-}, E_{-}, \tau u_{-},\tau e_{-} A_{-})$, where 
	$V_{-} = \{u | (\tau u(u) \cap \mathcal{T}_1 \neq \emptyset) \textit{ and } ((\tau u(u) \cap \mathcal{T}_2 = \emptyset) \textit{ or } (\exists (u,v) \in E_{-})\}$, 
	$E_{-} = \{u | (\tau e(e) \cap \mathcal{T}_1 \neq \emptyset) \textit{ and } (\tau e(e) \cap \mathcal{T}_2 = \emptyset)  \}$, $\forall u \in V_{-}$,  $\tau u_{-}(u)=\tau u(u) \cap \mathcal{T}_1$, $\forall e \in E_{-}$, $\tau e_{-}=\tau e(e) \cap \mathcal{T}_1$ and $A_{-}$ includes all $A(u, t)$ for which $u \in V_{-}$ and $t \in  \tau u_{-}(u) \cup \tau e_{-}(u,v)$.
\end{itemize}
 Figure \ref{fig:335} depicts the union graph $G_{\cup}$ of the temporal graph of Fig. \ref{fig:1} for $\mathcal{T}_1=[t_0,t_0]$ and $\mathcal{T}_2=[t_1,t_1]$.

\begin{figure*}
\begin{subfigure}[b]{0.32\textwidth}
\includegraphics[width=\textwidth]{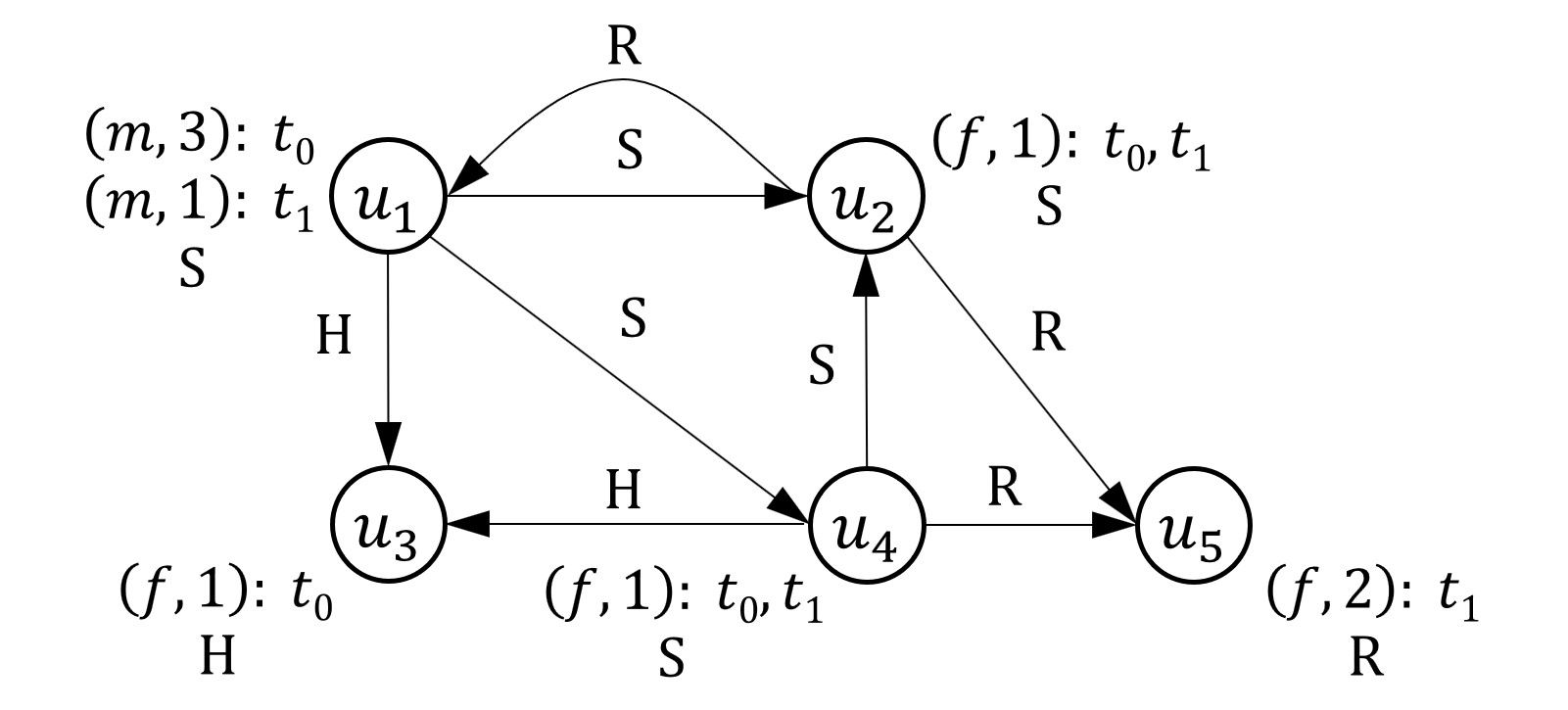}
\caption{}
\label{fig:sub3.1}
\end{subfigure}
\begin{subfigure}[b]{0.36\textwidth}
\includegraphics[width=\textwidth]{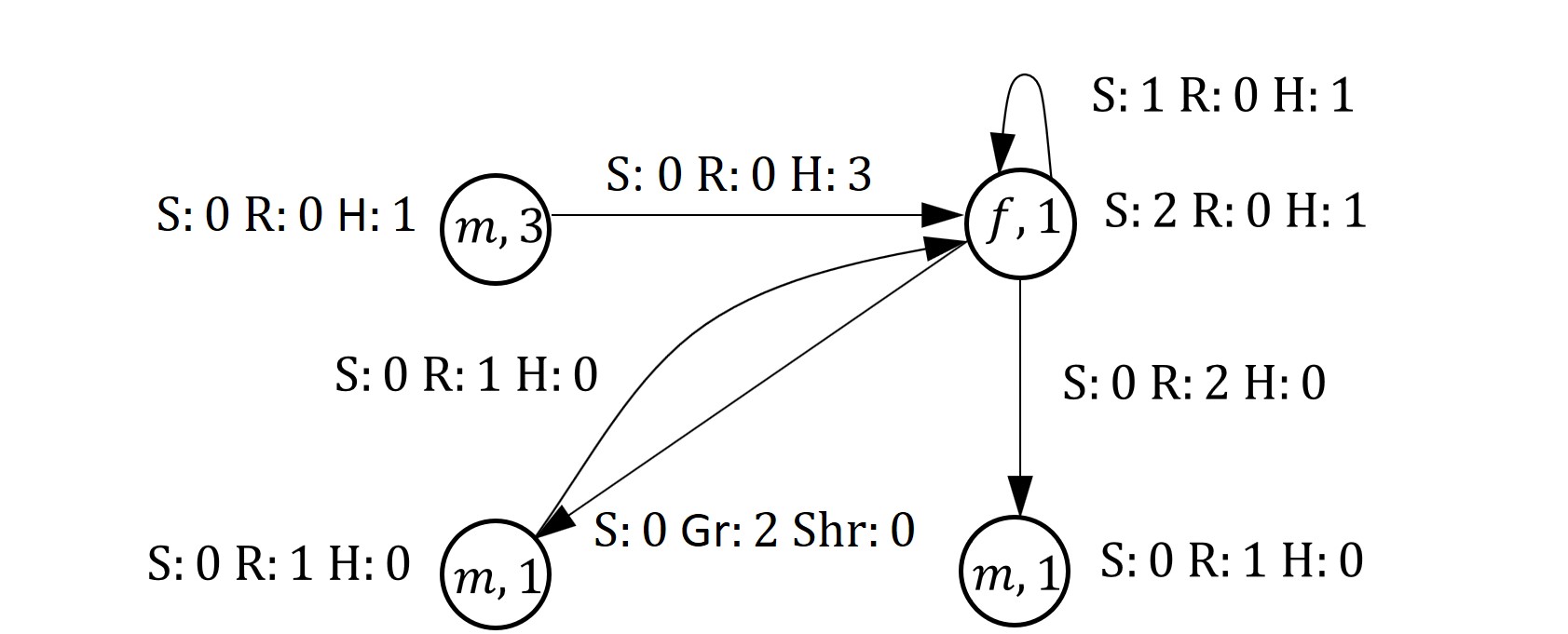}
\caption{}
\label{fig:sub3.2}
\end{subfigure}
\begin{subfigure}[b]{0.32\textwidth}
\includegraphics[width=\textwidth]{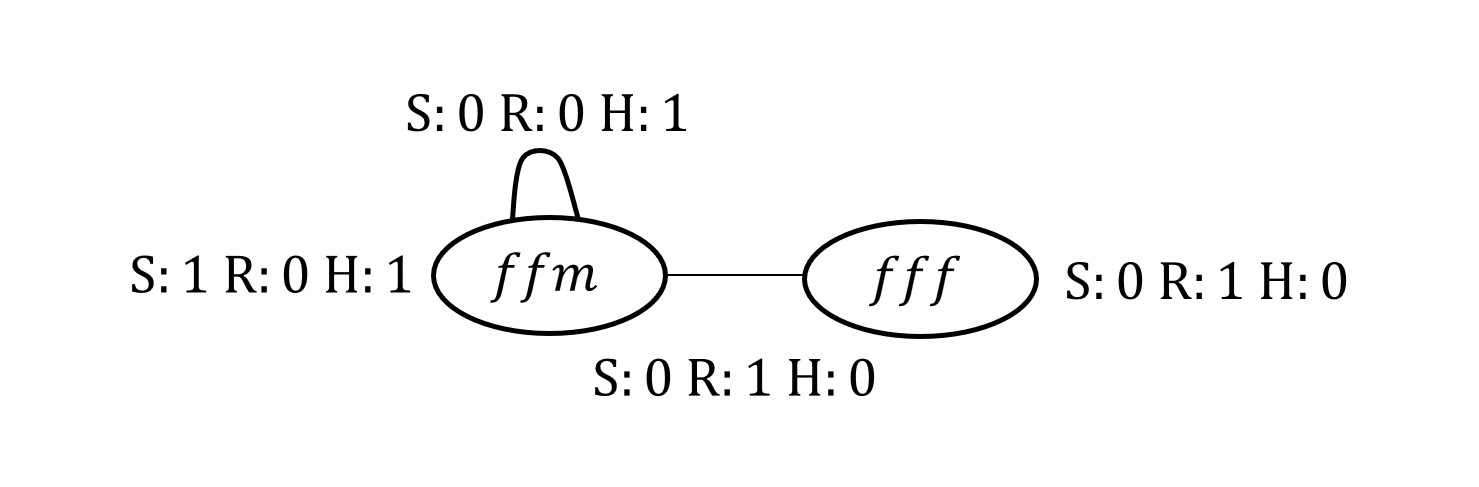}
\caption{}
\label{fig:sub3.3}
\end{subfigure}
\caption{(a) Evolution graph on (Gender, \#Publications), (b) its aggregation, and (c) triangle-based evolution graph aggregation on Gender for $t_0$, $t_1$ of the graph of Fig. \ref{fig:1}.}
\label{fig:3}
\end{figure*}

Temporal aggregation also enables us to study the changes that occur in a graph through time. Given $G$ and two sets of time intervals $\mathcal{T}_1$ and $\mathcal{T}_2$, its \textit{evolution} from $\mathcal{T}_1$ to $\mathcal{T}_2$ is captured by the following three graphs:
\begin{itemize}
    \item $G[\mathcal{T}_1 \cap \mathcal{T}_2]$ with entities that exist in both $\mathcal{T}_1$ and $\mathcal{T}_2$ and capture \textit{stability},
\item $G[\mathcal{T}_1 -\mathcal{T}_2]$, with entities that exist in $\mathcal{T}_1$ but not $\mathcal{T}_2$ and capture \textit{shrinkage}, and
\item $G[\mathcal{T}_2 -\mathcal{T}_1]$, with entities that did not exist in interval $\mathcal{T}_1$ but appear in $\mathcal{T}_2$ and capture \textit{growth}.
\end{itemize}
Overlaying the three graphs, we represent them as one \textit{evolution graph} as shown in Fig. \ref{fig:sub3.1} depicting the evolution of the graph of Fig.\ref{fig:1} from  $[t_0, t_0]$ to $[t_1, t_1]$, where labels $S$, $G$ and $R$ differentiate between entities with stability, growth and shrinkage respectively.

\begin{figure*}
\centering
\begin{subfigure}[b]{0.15\textwidth}
\includegraphics[width=\textwidth]{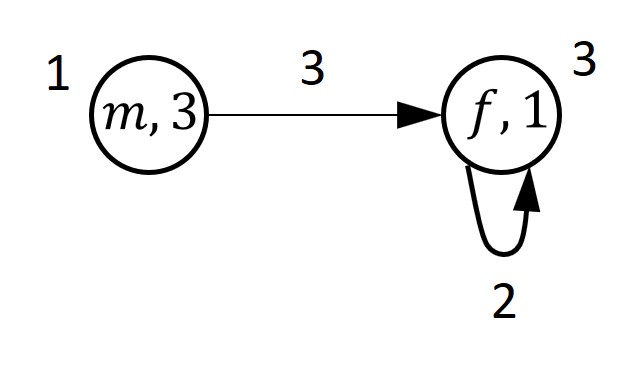}
\caption{\textit{$t_0$}}
\label{fig:sub2.1}
\end{subfigure}\hspace{5mm}%
\begin{subfigure}[b]{0.15\textwidth}
\includegraphics[width=\textwidth]{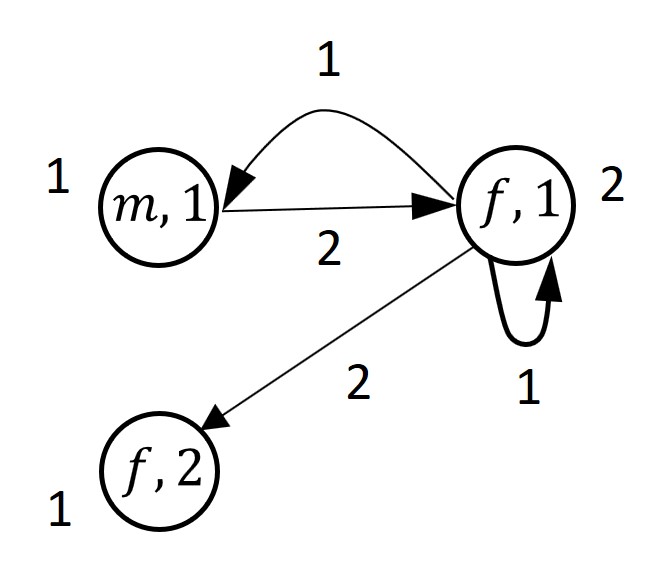}
\caption{\textit{$t_1$}}
\label{fig:sub2.2}
\end{subfigure}\hspace{5mm}%
\begin{subfigure}[b]{0.15\textwidth}
\includegraphics[width=\textwidth]{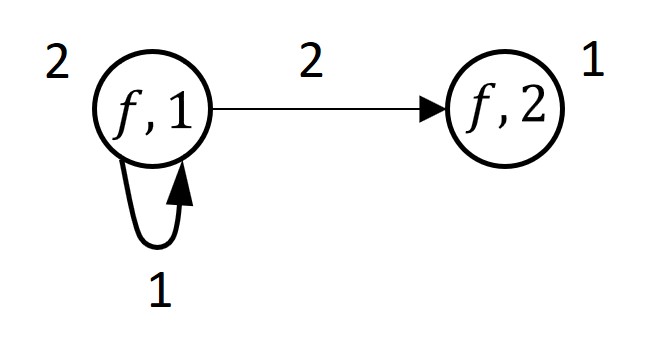}
\caption{\textit{$t_2$}}
\label{fig:sub2.3}
\end{subfigure}\hspace{5mm}%
\begin{subfigure}[b]{0.15\textwidth}
\includegraphics[width=\textwidth]{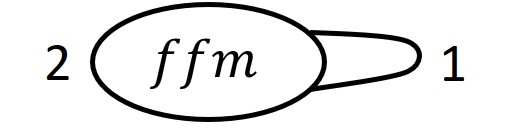}
\caption{\textit{$t_0$}}
\label{fig:sub2.4}
\end{subfigure}\hspace{5mm}%
\begin{subfigure}[b]{0.2\textwidth}
\includegraphics[width=\textwidth]{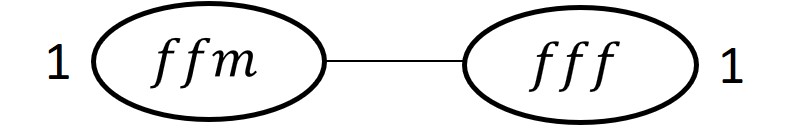}
\caption{\textit{$t_1$}}
\label{fig:sub2.5}
\end{subfigure}\hspace{5mm}%
\caption{Time point aggregate (a-c) graphs on (gender, \#publications) and (d-e) tri-graphs on gender for the graph of Fig.\ref{fig:1}.}
\label{fig:2}
\end{figure*}

\subsection{Graph Aggregation}
Besides temporal aggregation, we define graph or attribute aggregation by grouping nodes based on (some of) their attribute values, while taking network structure into account. So, each aggregate node (node in the aggregate graph) corresponds to a set of the original nodes, and aggregate edges are built based on the interactions of the aggregate nodes in the original graph. 

\begin{definition} [Graph Aggregation]
Given a temporal attributed graph $G(V, E, \tau u, \tau e, A)$ in $\mathcal{T}$, and $n$ aggregate attributes, $1\leq n \leq k$, $A'_1,A'_2,..,A'_n \in A$, the aggregated graph $G'$ is defined as a weighted graph $G'(V', E', w_{V'}, w_{E'}, A')$, where for each distinct tuple ($A^1(u,t), A^2(u,t),\dots,A^n(u,t)$), $u \in V$ and $t \in \tau u(u)$, there is a node $u' \in V'$ with $A'^i(u')=A^i(u,t), \forall i, 1 \leq i \leq n $, 
 and there is an edge $e' \in E'$ between $u', v' \in V'$, if and only if $\exists e \in E$ between nodes  $u,v \in V$ with $A'^i(u')=A^i(u,t)$  and $A'^i(v')=A^i(v,t), \forall i, 1 \leq i \leq n$. 
 The weights $w_{V'}$ and $w_{E}$ are aggregate functions $f_{V}$ and $f_{E}$ defined upon nodes and edges respectively.
\end{definition}
We use COUNT as our aggregation function for both nodes and edges, in the rest of this paper, and discern between \textit{distinct} aggregation (denoted as DIST) and \textit{non-distinct} aggregation (denoted as ALL). Distinct aggregation counts unique nodes and edges of the original graph, while, in non-distinct aggregation, duplicates are not identified and counted each time they appear.

\begin{figure*}
\centering
\begin{subfigure}[b]{0.15\textwidth}
\includegraphics[width=\textwidth]{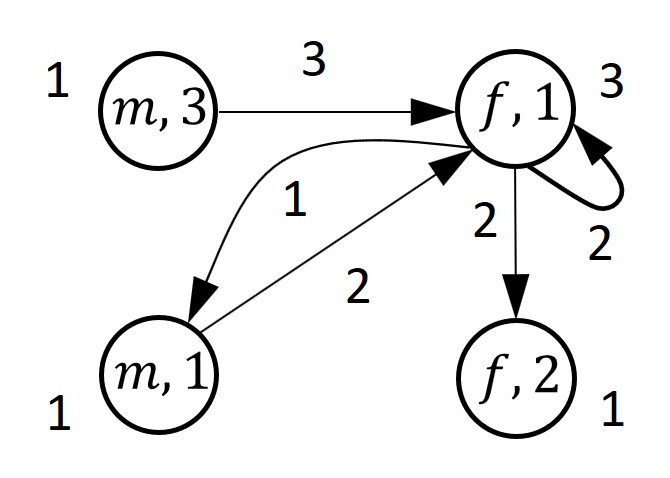}
\caption{\textit{$G'_{DIST}$}}
\label{fig:3tri.1}
\end{subfigure}\hspace{5mm}%
\begin{subfigure}[b]{0.15\textwidth}
\includegraphics[width=\textwidth]{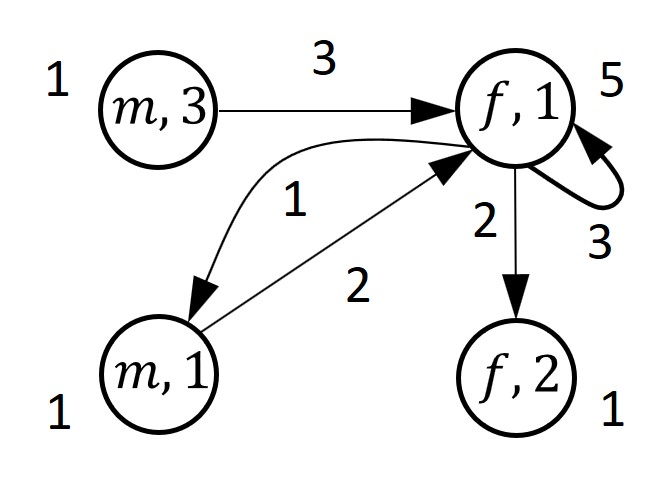}
\caption{\textit{$G'_{ALL}$}}
\label{fig:3tri.2}
\end{subfigure}\hspace{5mm}%
\begin{subfigure}[b]{0.2\textwidth}
\includegraphics[width=\textwidth]{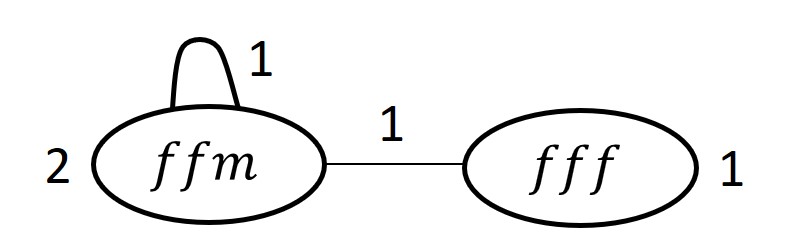}
\caption{\textit{$G'_{DIST}$}}
\label{fig:3tri.3}
\end{subfigure}\hspace{5mm}%
\begin{subfigure}[b]{0.2\textwidth}
\includegraphics[width=\textwidth]{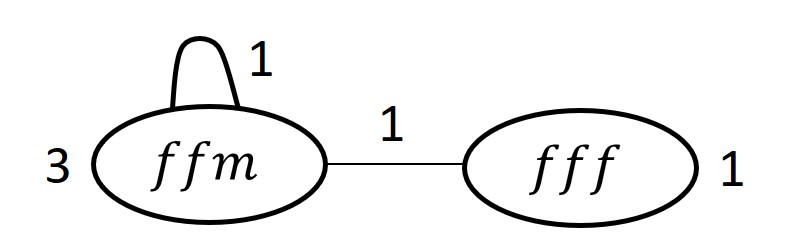}
\caption{\textit{$G'_{ALL}$}}
\label{fig:3tri.4}
\end{subfigure}\hspace{5mm}%
\caption{Aggregate (a-b) graphs on (gender, \#publications) and (c-d) triangle-based graphs on gender for the graph of Fig. \ref{fig:335}.}
\label{fig:3tri}
\end{figure*}

Figure \ref{fig:2}(a-c) depicts the aggregate graphs on gender and publications of the projections of the graph of Fig. \ref{fig:1} in
$t_0, t_1$ and $t_2$ respectively. As we show aggregate graphs on time points, there is no difference between distinct and non-distinct aggregation. We can see the difference in Fig. \ref{fig:3tri}(a-b) that shows  $G'_{DIST}$ (Fig. \ref{fig:3tri.1}) and $G'_{ALL}$  (Fig. \ref{fig:3tri.2}) for the union graph of Fig. \ref{fig:335}. The weight for node \textit{`f,1'}, of female authors with 1 publication, in $G'_{DIST}$ is 3, as 3 distinct nodes appear with this attribute value combination in the union graph, while in $G'_{ALL}$ it is 5, as there are 5 total appearances of this combination in the union graph.

Aggregation can be applied not only by grouping single nodes together, but subgraphs that match a specific pattern as well. For instance, we may be interested in closed triangles (i.e., cliques of size 3), and group them based on the attribute values of the nodes that form them. 

Let us define a \textit{pattern} $P$, as a non-attributed weakly connected subgraph, and let $p$ denote the number of nodes forming the subgraph. Graph aggregation can then be defined on patterns so that all subgraphs in $G$ that match $P$ (i.e. have the same structure) are grouped based on the attribute values of the nodes that belong to them, so that for $n$ aggregate attributes, each group of subgraphs corresponds to a distinct combination of $p*n$ attribute values.

\begin{definition} [Pattern Aggregation]
Given a temporal attributed graph $G(V, E, \tau u, \tau e, A)$ in $\mathcal{T}$, structural pattern $P$ and $n$ aggregation attributes, $1\leq n \leq k$, the aggregated graph $G'_P$ is defined as an undirected weighted graph $G'_P(V', E', w_{V'}, w_{E'}, A')$, where there is a node $v' \in V'$ for each group of subgraphs matching $P$ in the original graph whose nodes have the same $p*n$ attribute values forming $A'(v')$, and there is an edge $e' \in E'$ between $u', v' \in V'$, if and only if any subgraphs belonging to the groups of $u'$ and $v'$ in $G$ share at least one common node.
 The weights $w_{V'}$ and $w_{E'}$ are aggregate functions defined on the patterns and their common nodes in $G$.
\end{definition}
We use COUNT as our aggregation function, similarly to graph aggregation. 

Figure \ref{fig:2}(d-e) shows the aggregate graphs when aggregating based on patterns consisting of undirected triangles and gender. In $t_0$, the aggregate graph consists of a single node \textit{`ffm'} that corresponds to triangles formed by 2 female and 1 male author, since in the graph of Fig. \ref{fig:1}, there are two such triangles $u_1,u_2,u_4$ and  $u_1, u_3, u_4$. As the two triangles share node $u_1$ there is a self-loop in the aggregate graph. In $t_1$, besides the \textit{`ffm'} node that corresponds to $u_1,u_2,u_4$, there is also a \textit{`fff'} node corresponding to  $u_2,u_4,u_5$, while the aggregate graph for $t_2$ is empty, as no triangles are formed in $t_2$ in the original graph. In Fig. \ref{fig:3tri}(c-d), we can see the same pattern aggregation for the union graph of Fig. \ref{fig:335}.  In Fig. \ref{fig:3tri}(c), the weight of node \textit{`ffm'} is 2 as there are 2 distinct triangles formed in the union graph, while in Fig. \ref{fig:3tri}(d) it is 3 as $u_1, u_2, u_4$ appears in both time points.


As evolution is captured by three temporal attributed graphs, aggregation may be applied as well to provide a concrete view of graph evolution. The three graphs representing graph evolution are aggregated and overlaid, while different weights corresponding to each entity are used so as to discern between growth, shrinkage and stability. Figure \ref{fig:sub3.2} depicts the aggregation of the evolution graph of Fig. \ref{fig:sub3.1}, where we can discern the weights for stability, growth and shrinkage. For instance, node \textit{`f,1'} has: a) stability weight 1, as it had 1 stable appearance on node $u_2$ at both $t_0$ and $t_1$, b) growth weight 0, because no new appearance of $(f,1)$ occurs and c) shrinkage weight 1, because one appearance on $u_3$ at $t_0$ is removed in $t_1$. Similarly, Fig. \ref{fig:sub3.3} shows the triangle-based aggregate evolution graph of Fig. \ref{fig:sub3.1}, where for instance node \textit{`ffm'} has stability weight 1, as $u_2,u_1,u_4$ remains at both $t_0$ and $t_1$, growth weight 0, and shrinkage weight 1 as $u_1,u_4,u_3$ at $t_0$ disappears in $t_1$.
\section{Exploration}
We will now explore the evolution graph to detect interesting events through time. We are interested in three event types, namely \textit{growth}, \textit{shrinkage} and \textit{stability} which correspond respectively to  additions, deletions or persistence of graph elements, e.g., edges, nodes or patterns and their connections. We want to study different time granularities and different aggregation attributes both on the individual node and pattern level.

\subsection{Union and Intersection Semantics}
We study the evolution of the graph between two sets of time intervals ($\mathcal{T}_i$, $\mathcal{T}_j$) by studying the corresponding aggregate graphs of these intervals. Our goal is to detect interesting interval pairs such that many elements have been added (growth), deleted (shrinkage) or remained (stability) between $\mathcal{T}_i$ and $\mathcal{T}_j$.

Since the number of potential sets of intervals and corresponding aggregate graphs is exponential, we need a way to efficiently explore the corresponding search space. We start from intervals consisting of single time points, i.e., $\mathcal{T}_i=[t_i,t_i]$. If we consider $\mathcal{T}_i$ as the elements of a set $\mathcal{T}$, we
may construct the lattice of the powerset $\mathcal{T}$. Since we want to extend intervals based on combining successive intervals, we only focus on a sub lattice of the powerset lattice of $\mathcal{T}$, that is the tree that combines at each level $t$ of the lattice two elements of the previous level $t-1$ if they differ only by one element. As a powerset lattice is defined on union and intersection set operators, it follows by the definitions of the temporal operators union and intersection, that the graphs defined on the time intervals of the sub lattice also form a sub lattice with respect to these operations. 

\begin{definition}[Monotonically Increasing/Decreasing]
Given intervals $\mathcal{T}_k, \mathcal{T}_i, \mathcal{T}_j$ such that $\mathcal{T}_i \subseteq \mathcal{T}_j$ and aggregate graphs $G_i(V_i,E_i, W_{V_i}, W_{E_i}, A_i)$ and $G_j(V_j,E_j, W_{V_j}, W_{E_j}, A_j)$ defined as $G_i[\mathcal{T}_k \cdot \mathcal{T}_i]$ and $G_j[\mathcal{T}_k \cdot \mathcal{T}_j]$ respectively, temporal aggregation w.r.t operator $(\cdot)$, is defined as: 
\begin{itemize}
    \item \textbf{monotonically increasing} or \textbf{increasing} if it holds that (i) $\forall v \in V_j$, if $v \in V_i$ then $w_{V_i}(v) \leq w_{V_j}(v)$, and  (ii) $\forall e \in E_j$, if $e \in E_i$ then $w_{E_i}(e) \leq w_{E_j}(e)$, and
    \item \textbf{monotonically decreasing} or \textbf{decreasing} if it holds that (i) $\forall v \in V_j$, if $v \in V_i$ then $w_{V_i}(v) \geq w_{V_j}(v)$, and  (ii) $\forall e \in E_j$, if $e \in E_i$ then $w_{E_i}(e) \geq w_{E_j}(e)$.
\end{itemize}
\end{definition}
Regarding union and intersection semantics it easily follows from their definition that:
\begin{lemma}
Temporal aggregation is monotonically increasing with union and monotonically decreasing with intersection.
\end{lemma}

Thus, expanding intervals with union increases the size of the aggregate graphs and the number of events we find, while expanding with intersection reduces both the size of the graph and the event. Thus, the highest number of events of interest are detected when considering the largest intervals with union, and the smallest with intersection. Consequently, it is interesting to detect the minimal or maximal interval pairs at which events occur depending on the use of union or intersection respectively.
\begin{definition}[Minimal/Maximal Interval Pair]
An interval pair $\mathcal{T}_i, \mathcal{T}_j$ for which a number $k$ of events occurred from $\mathcal{T}_i$ to $\mathcal{T}_j$, is defined as: \begin{itemize}
    \item \textbf{minimal} if for the same $\mathcal{T}_i, \nexists \mathcal{T}_{j'} \subset \mathcal{T}_j$ for which $k'\geq k$ events occurred from $\mathcal{T}_i$ to $\mathcal{T}_{j'}$, and
    \item \textbf{maximal} if for the same $\mathcal{T}_i, \nexists \mathcal{T}_{j'}: \mathcal{T}_j \subset \mathcal{T}_{j'}$ for which $k'\geq k$ events occurred from $\mathcal{T}_i$ to $\mathcal{T}_{j'}$.
\end{itemize} 
\end{definition}

Based on the above, we define our problem as follows.
\begin{definition}[Problem Definition]
Given an attributed graph $G$, a time period $\mathcal{T}$ and a user-defined threshold $k$, find the minimal and maximal interval pairs in which at least $k$ events of either stability, growth or shrinkage occur.
\end{definition} 

\subsection{Stability}
Let us focus on the stability issue first.   Given two time intervals $\mathcal{T}_{old}$ and $\mathcal{T}_{new}$, where $\mathcal{T}_{old}$ precedes $\mathcal{T}_{new}$, we study stability by studying the aggregate graph $G[\mathcal{T}_{old}\cap\mathcal{T}_{new}]$.  The weights of the entities in the aggregate graph are the corresponding numbers of entities between $\mathcal{T}_{old}$ and $\mathcal{T}_{new}$ that remained stable. 

We propose a more efficient exploration of the candidate interval pairs by exploiting their  monotonicity properties. 
We keep one of the two ends of the interval fixed, as a  reference, and since we start with single intervals as our starting points we refer to this as \textit{reference point}. We gradually extend the other end of the interval by exploiting  union or intersection. When our focus is on the original graph then we maintain $\mathcal{T}_{old}$ fixed as our reference point while when our focus is on the latest state of the graph, $\mathcal{T}_{new}$ is viewed as our fixed reference point. 

 We first consider union semantics, thus, our problem is defined as finding the minimal interval pairs in which at least $k$ entities remain stable.  

Without loss of generality, let $w$ be the weight in the aggregate graph of the entity we are interested in, and let us further assume that we maintain $\mathcal{T}_{old}$ fixed and gradually extend $\mathcal{T}_{new}$. 
Given $\mathcal{T}_i, 1\leq i\leq n$,  we apply the following steps:  
\begin{enumerate}
    \item First compute all aggregate graphs  $G[\mathcal{T}_i \cap \mathcal{T}_{i+1}]$, $1\leq i<n$.
    \item If for any of the aggregate graphs $G[\mathcal{T}_i \cap \mathcal{T}_{i+1}]$, it holds that $w\geq k$, return $G[\mathcal{T}_i \cap \mathcal{T}_{i+1}]$ and prune any pair with $\mathcal{T}_i$ as its left end.
    \item For each pair  $\mathcal{T}_i, \mathcal{T}_{i+1}$ that continues, extend its right end by substituting $\mathcal{T}_{i+1}$ with its right child in the union semi-lattice (i.e., $\mathcal{T}_{i+1} \cup \mathcal{T}_{i+2}$), and apply step 2.
    \item The previous is repeated by extending the right end of each surviving interval with its right child in the union semi-lattice (i.e., $\mathcal{T}_{i+1} \cup \mathcal{T}_{i+2}\cup \dots \cup \mathcal{T}_{i+l}$) until either $i+l=n$ or no interval survives further.
\end{enumerate}
We refer to this algorithm as \textit{Union Exploration, (U-Explore)}.
U-Explore can be adapted to perform extension on the left end of each interval, i.e., by maintaining $\mathcal{T}_{new}$ fixed, and extending $\mathcal{T}_{old}$ by substituting in the next step by its left child in the union semi-lattice. 
\begin{theorem}
The minimal interval pairs for stability derived by extending $\mathcal{T}_{new}$ are not equal to those derived by extending $\mathcal{T}_{old}$. 
\end{theorem}
Theorem and lemma proofs can be found in \cite{Tsoukanara:supp}.

Let us now focus on intersection semantics, thus, our problem is finding the maximal intervals pairs in which at least $k$ nodes remain stable. Without loss of generality, let as assume that we maintain $\mathcal{T}_{old}$ and gradually extend $\mathcal{T}_{new}$.
Given $\mathcal{T}_i, 1\leq i\leq n$,  we apply the following steps:  
\begin{enumerate}
    \item First compute all aggregate graphs  $G[\mathcal{T}_i \cap \mathcal{T}_{i+1}]$, $1\leq i<n$.
    \item If for any aggregate graph $G[\mathcal{T}_i \cap \mathcal{T}_{i+1}]$, it holds that $w\geq k$, add $G$ to candidate set $C$, else prune  any pair with $\mathcal{T}_i$ as its left end. 
    \item For each pair  $\mathcal{T}_i, \mathcal{T}_{i+1}$ that continues, extend its right end by substituting $\mathcal{T}_{i+1}$ with its right child in the intersection semi-lattice (i.e., $\mathcal{T}_{i+1} \cap \mathcal{T}_{i+2}$)
    and apply step 2, and for any graph $G$ added to $C$, remove from $C$ its predecessor (i.e., the aggregate graph $G[\mathcal{T}_i \cap \mathcal{T}_{i+1}]$). 
    \item The previous is repeated by extending the right end of each surviving interval with its right child in the intersection semi-lattice (i.e., $\mathcal{T}_{i+1} \cap \mathcal{T}_{i+2}\cap \dots \cap \mathcal{T}_{i+l}$) until either $i+l=n$ or no interval survives further.
    \item The final candidate set $C$ is returned.
\end{enumerate}
We refer to this algorithm as \textit{Intersection Exploration, (I-Explore)}.
I-Explore can be adapted to perform extension on the left end of each interval, i.e., by maintaining $\mathcal{T}_{new}$ fixed, and extending $\mathcal{T}_{old}$ by substituting it in the next step by its left child in the intersection semi-lattice.

\begin{theorem}
The maximal interval pairs for stability derived by extending $\mathcal{T}_{new}$ are  equivalent to those derived by extending $\mathcal{T}_{old}$. 
\end{theorem}
\begin{table*}
\centering
\caption{Exploration Intervals Properties}
\label{tab:expl}
\setlength\tabcolsep{2pt}
\begin{adjustbox}{width=1\textwidth}
\begin{tabular}{c|c|c|ccccc}
\hline
\centering
\ Event & Type & Case & Left & Right & Mon. Increasing & Mon. Decreasing & $\subseteq$ of\\ \hline
\multirow{4}{*}{Growth} & \multirow{2}{*}{Min.} & $\mathcal{T}_{new}$ - $\mathcal{T}_{old}$($\cup$) & t.p. & t.p. & & \checkmark & $\mathcal{T}_{new}$($\cup$) - $\mathcal{T}_{old}$\\
\ & & $\mathcal{T}_{new}$($\cup$) - $\mathcal{T}_{old}$ & t.p. / interval & t.p. & \checkmark & &\\
\ & \multirow{2}{*}{Max.} & $\mathcal{T}_{new}$ - $\mathcal{T}_{old}$($\cap$) & t.p. & longest interval & \checkmark & &\\
\ & & $\mathcal{T}_{new}$($\cap$) - $\mathcal{T}_{old}$ & t.p. / interval & t.p. & & \checkmark &\\ \hline\hline
\multirow{4}{*}{Shrinkage} & \multirow{2}{*}{Min.} & $\mathcal{T}_{old}$($\cup$) - $\mathcal{T}_{new}$ & t.p. / interval & t.p. & \checkmark & &\\
\ & & $\mathcal{T}_{old}$ - $\mathcal{T}_{new}$($\cup$) & t.p. & t.p. & & \checkmark & $\mathcal{T}_{old}$($\cup$) - $\mathcal{T}_{new}$\\
\ & \multirow{2}{*}{Max.} & $\mathcal{T}_{old}$($\cap$) - $\mathcal{T}_{new}$ & t.p. / interval & t.p. & & \checkmark &\\
\ & & $\mathcal{T}_{old}$ - $\mathcal{T}_{new}$($\cap$) & t.p. & longest interval & \checkmark & &\\ \hline\hline
\multirow{4}{*}{Stability} & \multirow{2}{*}{Min.} & $\mathcal{T}_{old}$($\cup$) $\cap$ $\mathcal{T}_{new}$ & t.p. / interval & t.p. & \checkmark & & \\
\ & & $\mathcal{T}_{new}$($\cup$) $\cap$ $\mathcal{T}_{old}$ & t.p. / interval & t.p. & \checkmark & &\\
\ & \multirow{2}{*}{Max.} & $\mathcal{T}_{old}$($\cap$) $\cap$ $\mathcal{T}_{new}$ & t.p. / interval & t.p. & & \checkmark & $\mathcal{T}_{new}$($\cap$) $\cap$ $\mathcal{T}_{old}$\\
\ & & $\mathcal{T}_{new}$($\cap$) $\cap$ $\mathcal{T}_{old}$ & t.p. / interval & t.p. & & \checkmark & $\mathcal{T}_{old}$($\cap$) $\cap$ $\mathcal{T}_{new}$\\ \hline
\end{tabular}
\end{adjustbox}
\end{table*}
\subsection{Growth and Shrinkage}
For growth, we study the aggregate graph $G[\mathcal{T}_{new}-\mathcal{T}_{old}]$.

With union semantics, our problem is defined as finding the minimal interval pairs in which at least $k$ entities are added. As difference is not a symmetric operator, when extending $\mathcal{T}_{old}$, we expect $w$ to decrease, while when extending  $\mathcal{T}_{new}$, we expect an increase to $w$. 

Let us first extend $\mathcal{T}_{old}$. We adopt U-Explore by computing at step (1), instead of the intersection, the difference graphs $G[\mathcal{T}_{i+1} - \mathcal{T}_i]$, and then apply step (2). We claim that we do not need to proceed to further steps as  for any aggregate graph for which $w<k$ extending the right interval will not increase the result size on the derived aggregate graph. 

\begin{lemma} Temporal aggregation with difference $G[\mathcal{T}_{new}-\mathcal{T}_{old}]$ is monotonically decreasing when we extend $\mathcal{T}_{old}$ and monotonically increasing when we extend $\mathcal{T}_{new}$, using union semantics.
\label{lem:anti-union}
\end{lemma}

Let us now consider extending $\mathcal{T}_{new}$. Again, we adopt U-Explore, by computing the difference aggregate graphs $G[\mathcal{T}_{i+1} - \mathcal{T}_i]$ instead of the intersection aggregate graphs. The rest of the steps are applied as defined. 

With intersection semantics, we determine the maximal intervals pairs in which at least $k$ entities are added. We first extend $\mathcal{T}_{old}$.
\begin{lemma}
Temporal aggregation with difference $G[\mathcal{T}_{new}-\mathcal{T}_{old}]$ is monotonically increasing when we extend $\mathcal{T}_{old}$ and monotonically decreasing when we extend $\mathcal{T}_{new}$, using intersection semantics.
\label{lem:mon-inter}
\end{lemma}

Based on the above lemma, the maximal interval pairs are: each point of reference  $\mathcal{T}_{new}$, with the longest possible $\mathcal{T}_{old}$ as long as for the defined aggregate graph, $w \geq k$.

When extending $\mathcal{T}_{new}$, I-Explore is adopted appropriately. In all respective steps, we compute the aggregate difference graphs $G[\mathcal{T}_{i+1} - \mathcal{T}_i]$, and then apply the rest of the steps accordingly.


To study shrinkage, we study $G[\mathcal{T}_{old}-\mathcal{T}_{new}]$ and our problem is symmetrical to growth. Thus, with union semantics we are interested in finding minimal interval pairs and U-Explore is deployed, while with intersection semantics we are interested in maximal interval pairs and use I-Explore.

All cases are summarized on Table \ref{tab:expl}. Column \textit{Case} shows the reference point and the extended interval using either union ($\cup$) or intersection ($\cap$) semantics. Columns \textit{Left} and \textit{Right} refer to the left and right interval in the result interval pairs. One of the elements in the pair is always a time point (t.p.) while the other can be either time point or interval and longest interval. The last column shows the relationships between the results of different cases, and in particular when the result of one case is a subset of the result of another case.

\subsection{Initialization of \textit{k}}
The appropriate value of the threshold $k$ depends each time on the given graph and its data. To configure this threshold, we propose an initial value for $k$, indicating a threshold of interestingness and acting as a starting point so as to attain some results and then continue with its tuning. In particular, given the type of entity we are interested in, we initialize $k$ to a value $w_{th}$ which is defined as the minimum or the maximum weight of the given type of entity in the aggregation graph for any graph defined on the temporal aggregation of pairs of consecutive time points. For stability, we compute the aggregate graphs on intersections of pairs of consecutive time points, while for growth and shrinkage we compute the aggregate graphs on the appropriate difference graphs. Further, for a monotonically increasing operator, we start with  $w_{th}$ as the minimum aggregation weight for a given entity type, and increase it gradually, while for the monotonically decreasing operators $w_{th}$ corresponds to the maximum aggregation weight and is gradually decreased as the exploration progresses.


\begin{table}
\caption{Arrays $\mathbf{V}$, $\mathbf{S}$ for Gender, and $\mathbf{A}$ for \#Publications for the graph of Fig. \ref{fig:1}.}
\label{tab:s1}
\parbox{.3\linewidth}{
\centering
\begin{adjustbox}{width=0.15\textwidth}
\begin{tabular}{|c|c|c|c|}
\hline
\textbf{{Id}}&{$t_0$}&{$t_1$}&{$t_2$}\\ \hline
\textbf{$u_1$} & {$1$} & {$1$} & {$0$}\\ \hline
\textbf{$u_2$} & {$1$} & {$1$} & {$1$}\\ \hline
\textbf{$u_3$} & {$1$} & {$0$} & {$0$}\\ \hline
\textbf{$u_4$} & {$1$} & {$1$} & {$1$}\\ \hline
\textbf{$u_5$} & {$0$} & {$1$} & {$1$}\\ \hline
\end{tabular}
\end{adjustbox}
}%
\parbox{.3\linewidth}{
\centering
\begin{adjustbox}{width=0.078\textwidth}
\begin{tabular}{|c|c|}
\hline
\textbf{Id}&{}\\ \hline
\textbf{$u_1$} & $m$\\ \hline
\textbf{$u_2$} & $f$\\ \hline
\textbf{$u_3$} & $f$\\ \hline
\textbf{$u_4$} & $f$\\ \hline
\textbf{$u_5$} & $f$\\ \hline
\end{tabular}
\end{adjustbox}
}
\parbox{.3\linewidth}{
\centering
\begin{adjustbox}{width=0.15\textwidth}
\begin{tabular}{|c|c|c|c|}
\hline
{\textbf{Id}}&{$t_0$}&{$t_1$}&{$t_2$}\\ \hline
\textbf{$u_1$} & {$3$} & {$1$} & {-}\\ \hline
\textbf{$u_2$} & {$1$} & {$1$} & {$1$}\\ \hline
\textbf{$u_3$} & {$1$} & {-} & {-}\\ \hline
\textbf{$u_4$} & {$2$} & {$1$} & {$1$}\\ \hline
\textbf{$u_5$} & {-} & {$2$} & {$2$}\\ \hline
\end{tabular}
\end{adjustbox}
}%
\end{table}

\section{Algorithms}
\label{sec:tcube}

To store an attributed temporal graph $G(V, E, \tau u, \tau e, A)$ defined in a set of time intervals $\mathcal{T}$, we maintain separate structures for $V$, $E$ and $A$. For each $v \in V$,  $\tau u(u)$ is represented by a binary vector of size $|\mathcal{T}|$, where each element in the vector corresponds to a time $t\in \mathcal{T}$, and is 1 iff $t \in \tau u(v)$. Combining the temporal information for all nodes, we store these vectors as a labeled array $\mathbf{V}$ with $|V|$ rows labeled with the nodes ids and $|\mathcal{T}|$ columns labeled with the time $t$. A similar representation is deployed for the edges that are also stored in a labeled array $\mathbf{E}$, with $|E|$ rows labeled with the edges ids denoted as the edges end points $(u,v)$ and  $|\mathcal{T}|$ columns labeled with the time $t$. 

For efficient storage and processing, we discern between static and time-varying attributes. For static-attributes a labeled array $\mathbf{S}$ maintains a row for each node $v \in V$, labeled by the node id, and a number of columns equal to the number of static attributes, labeled by the corresponding attribute name. Each row $v$ in the array associates node $v$ with its corresponding static attributes values. In contrast, we maintain a labeled array $\mathbf{A}^i$, for each of the time-varying attributes $A^i$. Rows are similarly to $\mathbf{S}$ labeled with node ids, while the columns maintain temporal information. Each cell $\mathbf{A}^i[v,t]$ maintains the value of attribute $A^i$ of $v$ at time $t$. Table \ref{tab:s1} depicts $\mathbf{V}$,  $\mathbf{S}$ and $\mathbf{A}$ for the  graph of Fig. \ref{fig:1}. 

\subsection{Temporal Aggregation Implementation}
Based on our data representation, time projection is easily implemented by restricting the arrays to the columns corresponding to a given time interval.

We present in detail the algorithm implementing the union operator 
 of a graph $G(V, E, A)$ for $\mathcal{T}_1, \mathcal{T}_2$ as depicted in Alg. \ref{algo:Union}. The algorithm takes as input the labeled arrays $\mathbf{V}, \mathbf{E}, \mathbf{S} \text{ and } \mathbf{A}^i$ for all time-varying attributes $A^i \in A$, and outputs the labeled arrays $\mathbf{V}_\cup, \mathbf{E}_\cup, \mathbf{S}_\cup \text{ and } \mathbf{A}_{i\cup}$ that maintain the information of the derived union graph $G[\mathcal{T}_1 \cup \mathcal{T}_2]$. 
As our result is defined in $(\mathcal{T}_1, \mathcal{T}_2)$, we initialize all $\mathbf{V}_\cup, \mathbf{E}_\cup$ and $\mathbf{A}^i_{\cup}$s as empty arrays with one column for each $t \in  \mathcal{T}_1 \cup \mathcal{T}_2$ (line 1). 
Similarly, we restrict the input tables, except $\mathbf{S}$ that stores no temporal information, only to the columns corresponding to time $t \in \mathcal{T}_1 \cup \mathcal{T}_2$ (line 2). 
 For each row $v$ of $\mathbf{V}$, if there is a value in any $\mathbf{V}[v,t]$ equal to $1$, the row is inserted in $\mathbf{V}_\cup$. The corresponding entries for the attributes of $v$ are updated. That is, row $v$ in $\mathbf{S}$ and for each time-varying attribute $A^i$ the corresponding rows from arrays $\mathbf{A}^{i}$ are copied to $\mathbf{S}_\cup$ and $\mathbf{A}_{i\cup}$ respectively (lines 3-7). Similarly, for each row $e$ of $\mathbf{E}$, if there is a value in any $\mathbf{E}[e,t]$ equal to $1$, the row is inserted in $\mathbf{E}_\cup$ (lines 8-10).

Intersection is applied similarly to union, only for a row $v$ of $\mathbf{V}$ to be inserted into $\mathbf{V}_\cap$, we require that all elements $\mathbf{V}[v,t]$ with $t \in \mathcal{T}_1$ and $\mathbf{V}[v,t']$ with $t' \in \mathcal{T}_2$ are equal to 1. The insertion of edges is also changed accordingly.

Finally, for the difference operator, let $G[\mathcal{T}_1-\mathcal{T}_2]$, a row $v$ of $\mathbf{V}$  is inserted into $\mathbf{V}_{-}$ if any $\mathbf{V}[v,t]$ with $t \in \mathcal{T}_1$ is equal to 1, and all $\mathbf{V}[v,t']$ with $t' \in \mathcal{T}_2$ are equal to 0.

\subsection{Attribute Aggregation Implementation}
Aggregation differentiates between distinct and non-distinct. Furthermore, to improve efficiency, static and time-varying attributes can be treated separately. 

Algorithm \ref{algo:AggVarDist} applies distinct aggregation when at least one of the attributes of aggregation is time-varying. It takes as input  $\mathbf{V}, \mathbf{E}, \mathbf{S} \text{ and } \mathbf{A}^i$ for all time-varying attributes $A^i \in A$, interval $\mathcal{T}$ and a list $L$ of the aggregation attributes, and outputs the labeled vectors $\mathbf{V'}, \mathbf{E'}$ representing the aggregated graph $G'$.
For each attribute $A^i$ in $L$, the array $\mathbf{A}^i$ is unpivoted, i.e., its columns corresponding to time points in $\mathcal{T}$ are appended as rows (lines 1-2), and then merged into a new array  $\mathbf{A'}$ (line 3). $\mathbf{A'}$ is then deduplicated based on key $(u, a')$, where $a'$ is a tuple formed by the aggregate attributes, so that each node $u$ appears with a given attribute tuple $a'$ only once (line 4). For static attributes, we restrict $\mathbf{S}$ to the columns corresponding to attributes in $L$ (line 5) and merge them into $\mathbf{A'}$ (line 6). Groups are then formed on $\mathbf{A'}$ based on each distinct attribute tuple $a'$ and its appearances are counted (line 7). Each such group corresponds to a node that is inserted in the aggregate graph, labeled by the attribute tuple $a'$ and with value equal to the groups count which constitutes the node's weight (lines 8-10). To add the edges in the aggregation graph, we traverse the edge array $\mathbf{E}$ and lookup the edge ends (nodes) in $\mathbf{A'}$ to retrieve the pair of attribute tuples corresponding to the given nodes at each time point. Each such tuple pair is inserted into a new temporary array $\mathbf{A''}$ (lines 11-14). The rest of the procedure is similar to our treatment of nodes.  $\mathbf{A''}$ is deduplicated based on key $(u,v),(a',a'')$, groups are formed based on pairs $(a',a'')$ and their appearances counted (lines 15-16). Finally, for each group a new row with label $(a',a'')$ is inserted in $\mathbf{E'}$, representing a new edge,  and with weight equal to its count that is assigned as its value (lines 17-19). 

For non-distinct aggregation, Alg. \ref{algo:AggVarDist} is modified by omitting deduplication for nodes and edges, thus counting all appearances of each attribute tuple, or pair of attribute tuples respectively. 
 
Optimization is possible if aggregation concerns only static attributes. For distinct aggregation, we do not require unpivoting nor deduplication as each node has a unique value for each static attribute . Further, for each edge the lookups required for its ends do not depend on time and again no deduplication is required. For non-distinct aggregation, as we need to count all appearances of a node, we initialize corresponding weights for each $(u,a')$ (or $(u,v),(a',a'')$) by counting the columns in $\mathbf{V}$ (or $\mathbf{E}$) equal to 1. Then, instead of counting the appearances of each group $a'$ (or $(a',a'')$, we sum their weights. 

\begin{algorithm}
	\KwIn{A temporal attributed graph $G$ represented by $\mathbf{V}, \mathbf{E}, \mathbf{S} \text{ and } \{\mathbf{A}_i\}$, intervals $\mathcal{T}_1,\mathcal{T}_2$}
	\KwOut{The union graph $G_{\cup}$ represented by $\mathbf{V}_\cup, \mathbf{E}_\cup, \mathbf{S_\cup}, \{\mathbf{A}_{i\cup}\}$}
	
	Initialize $\mathbf{V}_\cup, \mathbf{E}_\cup, \{\mathbf{A}_{i\cup}\}$ in  $\mathcal{T}_1 \cup \mathcal{T}_2$, and $\mathbf{S_\cup}$\\
	Restrict  $\mathbf{V}, \mathbf{E}, \text{ and } \{\mathbf{A}_i\}$ in  $\mathcal{T}_1 \cup \mathcal{T}_2$\\
	
	\For{each row $v \in \mathbf{V}$}{ 
		\If(\tcp*[h]{$t \in \mathcal{T}_1 \cup \mathcal{T}_2$}){$any$ $\mathbf{V}[v,t] = 1$}{		
			Insert  $\mathbf{V}[v]$ in $\mathbf{V}_\cup$\\
			Insert $\mathbf{S}[v]$ in $\mathbf{S}_{\cup}$\\
			Insert each $\mathbf{A}_i[v]$ in the corresponding $\mathbf{A}_{i\cup}$\\ 
		}
	}
	\For{each row $e \in \mathbf{E}$} {
	\If(\tcp*[h]{$t \in \mathcal{T}_1 \cup \mathcal{T}_2$}){$any$ $\mathbf{E}[e,t] = 1$} {
		Insert $\mathbf{E}[e]$ in $\mathbf{E}_\cup$\\ 
	}
}
	\Return{$\mathbf{V}_\cup, \mathbf{E}_\cup, \mathbf{S_\cup}, \{\mathbf{A}_{i\cup}\}$}
	\caption{{\sc Union}}
	\label{algo:Union}
\end{algorithm}

\subsection{Pattern Aggregation Implementation}
For pattern-based aggregation, we need to group the nodes of the input temporal attributed graph based on the given aggregation pattern, and then count the connections between such groups. To facilitate this process, given a pattern $P$, we rely on an intermediate pattern-based temporal attributed graph, we call pattern graph, $G_P$, that has a node $v$ for each set of nodes that match pattern $P$ in $G$, and there is an edge $(u,v)$ in $G_P$ if there is at least one common node between the sets of nodes that correspond to $u$ and $v$. Here, we describe the algorithm for building the pattern graph when $P$ is an undirected closed triangle,  (Alg. \ref{algo:tri}). 

\begin{algorithm}
\KwIn{A temporal attributed graph $G$ represented by $\mathbf{V}, \mathbf{E}, \mathbf{S} \text{ and } \{\mathbf{A}_i\}$, intervals $\mathcal{T}$ and a list $L$ of aggregation attributes}
\KwOut{The aggregate graph $G'$ represented by arrays $\mathbf{V'}$ and $\mathbf{E'}$}
\For{$A_i \in L$} {
$\mathbf{A'}_i\leftarrow unpivot(\mathbf{A}_i)$}

{$\mathbf{A'}\leftarrow$ Merge all $\mathbf{A'}_i$s}

{$\mathbf{A'} \leftarrow \mathbf{A'}.deduplicate(v,a')$}

{$\mathbf{S'}$ restrict to columns for $s \in L$}

{$\mathbf{A'}\leftarrow$ Merge $\mathbf{A'}$ and $S$}

{$GroupsA'\leftarrow \mathbf{A'}.groupby(a').count()$}

\For{each group in $GroupsA'$} {
{Insert row $a'$ in $\mathbf{V'}$ }

{$\mathbf{V'}[a'] \leftarrow a'.count$}
}
\For{$(e(u, v),t) \in \mathbf{E}$} {
{$a' \leftarrow$ Look up $(u,t)$ in $\mathbf{A'}$}

{$a'' \leftarrow$ Look up $(v,t)$ in $\mathbf{A'}$}

{Insert $(a',a'')$ to $\mathbf{A''}$}
}
{$\mathbf{A''}.deduplicate((u,v),(a',a''))$}

{$GroupsA''\leftarrow \mathbf{A''}.groupby((a',a'')).count()$}

\For{each group in $GroupsA''$} {
{Insert row $(a',a'')$ to $\mathbf{E'}$}

{$\mathbf{E'}[(a',a'')] \leftarrow (a',a'').count$}
}
\Return{$\mathbf{V'},\mathbf{E'}$}
\caption{{\sc Distinct Aggregation}}
\label{algo:AggVarDist}
\end{algorithm}

 The algorithm takes as input the labeled arrays that constitute the graph $G$ in $\mathcal{T}$, that is, $\mathbf{V}$, $\mathbf{E}$, $\mathbf{S}$, and $\{\mathbf{A}_i\}$ and outputs the labeled arrays of the tri-graph ${G}_{tri}(\mathbf{V}_{tri}, \mathbf{E}_{tri}, \mathbf{S}_{tri}, \{\mathbf{A}_{{i}_{tri}}\})$ in $\mathcal{T}$. 
 First, we initialize $\mathbf{V}_{tri}$, $\mathbf{E}_{tri}$, $\mathbf{S}_{tri}$, $\mathbf{A}_{{i}_{tri}}$ as empty arrays (line 1). Next, we find all triangles that exist in $G$. 
 For each $v \in \mathbf{V}$, we locate all edges $(i,j)$ that close triangles between neighbors $i$ and $j$ of $v$ and add a triangle $(i, j, v)$ in $\mathbf{V}_{tri}$ with $1$ in column $t \in \mathcal{T}$ if $\mathbf{E}[(i,j),t]=\mathbf{E}[(i,v),t]=\mathbf{E}[(j,v),t]=1$ (lines 2-7).


To create the edges in the tri-graph, we use a hash table $\mathcal{H}$, where we store the node – triangles mappings (line 8). Specifically, each triangle $v_{tri}$ is inserted in $\mathcal{H}[v]$, $v \in v_{tri}$ (lines 9-10). 
For all pairs of triangles $(v_{tri}, n_{tri})$ in $\mathcal{H}[v]$, we set $\mathbf{E}_{tri}[(v_{tri}, n_{tri}), t]$ to 1, if both triangles exist in $t$ (lines 12-15).

Finally, $\mathbf{S}_{tri}, \{\mathbf{A}_{{i}_{tri}}\}$ are filled by retrieving for each node $v_{tri}=\{v_1,v_2,v_3\} \in \mathbf{V}_{tri}$ the corresponding rows of $v_1, v_2$ and $v_3$ in  $\mathbf{S}, \{\mathbf{A}_{{i}}\}$. For the time-varying attributes, the algorithm also needs to check the attribute values at each $t \in T$ (lines 16-20). 


We can utilize this pattern graph for pattern aggregation in two ways. In the first approach, we first create the tri-graph of the original temporal attributed graph and then apply any temporal operation and distinct or non-distinct aggregation on it. In the second approach, we first perform temporal aggregation on the original graph, and then construct the tri-graph of the result and apply pattern aggregation. The second approach, leverages a possibly smaller graph derived after applying the temporal operator, for operations such as intersection. 

\subsection{Optimizations}
To implement our aggregation framework, we need to materialize all possible combinations of dimensions and all possible interval aggregations. Although, this is the best option in terms of response time of OLAP queries, for a large multidimensional network it is quite unrealistic as it requires excessive storage space. We propose using partial materialization based on the properties of the aggregate graphs and which graphs can be derived by other aggregates without accessing the original graph. 

Given an aggregate graph $G'$ on a set of attributes $A'=\{A^1,A^2,\dots, A^k\}$ any aggregate graph $G''$ on $A'' \subseteq A'$ can be derived directly from $G'$. Nodes and edges between them in $G'$ with attributes corresponding to distinct tuples in $A''$ are grouped to form new aggregate nodes and edges in $G''$. Weights in $G''$ are evaluated by summing up the respective weights of the entities of $G'$ belonging to each group. Consequently,aggregation with COUNT is identified as a \textit{D-distributive} measure w.r.t. top-down aggregations.

With respect to time, we claim that union and non-distinct aggregation is \textit{T-distributive}. That is, to compute the weights of the higher level aggregate union graph, we sum up the weights for the respective nodes and edges in each of the lower level graphs. 
Based on the above, we propose precomputing aggregations 
 on the unit of time that can be combined with union to attain the aggregate graphs of higher granularity. However, distinct union aggregates are not T-distributive, as we need to identify distinct nodes across multiple graphs. 
\begin{algorithm}
	\KwIn{A temporal attributed graph ${G}$ represented by $\mathbf{V}, \mathbf{E}, \mathbf{S} \text{ and } \{\mathbf{A}_i\}$, set of intervals $\mathcal{T}$}
	\KwOut{The triangle-based graph $G_{tri}$ represented by $\mathbf{V}_{tri}, \mathbf{E}_{tri}, \mathbf{S}_{tri}, \{\mathbf{A}_{{i}_{tri}}\}$}
	Initialize $\mathbf{V}_{tri}, \mathbf{E}_{tri}, \mathbf{S}_{tri}, \{\mathbf{A}_{{i}_{tri}}\}$\\
    \For{each node $v$ in $\mathbf{V}$ and $t \in \mathcal{T}$}{
        Find all edges $(i, j) \in E$ for which $(i, v) \in E$ and $(j, v) \in E$\tcp{$i,j \neq v$}
        \If{$\mathbf{E}[(i, j), t] = 1$ \text{and} $\mathbf{E}[(i, v), t] = 1$ \text{and} $\mathbf{E}[(j, v), t] = 1$}
        {$\mathbf{V}_{tri}[(i, j, v), t] = 1$ }
        \Else{
            $\mathbf{V}_{tri}[(i, j, v), t] = 0$
        }
    }
    Initialize $\mathcal{H}$ hash table\\
    \For{each triangle $v_{tri}$ in $\mathbf{V}_{tri}$}{
        $\mathcal{H}[v_i] \gets v_{tri}$\tcp{ $v_i \in \mathcal{V}_i$} }
    \For{each pair $(v_{tri}, n_{tri})$ in $\mathcal{H}[v_i]$ and $t \in \mathcal{T}$}{
        \If{$\mathbf{V}_{tri}[v_{tri}, t] = 1$ and $\mathbf{V}_{tri}[n_{tri}, t] = 1$}{
            $\mathbf{E}_{tri}[(v_{tri}, n_{tri}), t] = 1$
        }
        \Else{
            $\mathbf{E}_{tri}[(v_{tri}, n_{tri}), t] = 0$
        }
    }
    \tcp{pseudocode for one time-varying attribute}
    \For{each triangle $(v_1, v_2, v_3)$ in 
 $\mathbf{V}_{tri}$ and $t \in \mathcal{T}$}{
        \If{$\mathbf{V}[v_1, t] = 1$ and $\mathbf{V}[v_2, t] = 1$ and $\mathbf{V}[v_3, t] = 1$}{
            $\mathbf{A}_{{1}_{tri}}[(v_1, v_2, v_3), t] = (\mathbf{A}_1[v_1, t], \mathbf{A}_1[v_2, t], \mathbf{A}_1[v_3, t])$ }
        \Else{
            $\mathbf{A}_{{1}_{tri}}[(v_1, v_2, v_3), t] = None$
        }
    }
    \tcp{$\mathbf{S}_{tri}$ is developed similarly omitting the dimension of time}
	\Return{$\mathbf{V}_{tri}, \mathbf{E}_{tri}, \mathbf{S}_{tri}, \{\mathbf{A}_{{i}_{tri}}\}$}
	\caption{{\sc Tri-Graph Construction}}
	\label{algo:tri}
\end{algorithm}

 \begin{table*}
\caption{\textit{DBLP} Graph}
\label{tab:dblp}
\centering
\begin{adjustbox}{width=1\textwidth}
\begin{tabular}{c|ccccccccccccccccccccc}
\hline
\centering
\textbf{\#TP} & 2000 & 2001 & 2002 & 2003 & 2004 & 2005 & 2006 & 2007 & 2008 & 2009 & 2010 & 2011 & 2012 & 2013 & 2014 & 2015 & 2016 & 2017 & 2018 & 2019 & 2020\\ \hline
\textbf{\#Nodes} & 1708 & 2165 & 1761 & 2827 & 3278 & 4466 & 4730 & 5193 & 5501 & 5363 & 6236 & 6535 & 6769 & 7457 & 7035 & 8581 & 8966 & 9660 & 11037 & 12377 & 12996\\
\textbf{\#Edges} & 2336 & 2949 & 2458 & 4130 & 4821 & 7145 & 7296 & 7620 & 8528 & 8740 & 10163 & 10090 & 11871 & 12989 & 12072 & 15844 & 16873 & 18470 & 21197 & 27455 & 28546\\
\hline
\end{tabular}
\end{adjustbox}
\end{table*}

\begin{table}
\caption{\textit{MovieLens} Graph}
\label{tab:movielens}
\centering
\begin{adjustbox}{width=.5\textwidth}
\begin{tabular}{c|cccccc}
\hline
\centering
\textbf{\#TP} & May & Jun & Jul & Aug & Sep & Oct\\ \hline
\textbf{\#Nodes} & 486 & 508 & 778 & 1309 & 575 & 498\\
\textbf{\#Edges} & 100202 & 85334 & 201800 & 610050 & 77216 & 48516\\
\hline
\end{tabular}
\end{adjustbox}
\end{table}

\begin{table*}
\caption{\textit{Primary School} Graph}
\label{tab:pm}
\centering
\begin{adjustbox}{width=1\textwidth}
\begin{tabular}{c|ccccccccccccccccc}
\hline
\centering
\textbf{\#TP} & 1 & 2 & 3 & 4 & 5 & 6 & 7 & 8 & 9 & 10 & 11 & 12 & 13 & 14 & 15 & 16 & 17\\ \hline
\textbf{\#Nodes} & 228 & 231 & 233 & 220 & 118 & 217 & 215 & 232 & 238 & 235 & 235 & 236 & 147 & 119 & 211 & 175 & 187\\
\textbf{\#Edges} & 857 & 2124 & 1765 & 1890 & 1253 & 1560 & 1051 & 1971 & 1170 & 1230 & 2039 & 1556 & 1654 & 1336 & 1457 & 1065 & 1767\\
\hline
\end{tabular}
\end{adjustbox}
\end{table*}

\begin{table*}
\caption{\textit{Primary School-tri} Graph}
\label{tab:pmtri}
\centering
\begin{adjustbox}{width=1\textwidth}
\begin{tabular}{c|ccccccccccccccccc}
\hline
\centering
\textbf{\#TP} & 1 & 2 & 3 & 4 & 5 & 6 & 7 & 8 & 9 & 10 & 11 & 12 & 13 & 14 & 15 & 16 & 17\\ \hline
\textbf{\#Nodes} & 1133 & 7231 & 4125 & 5388 & 3865 & 2785 & 2965 & 5815 & 1982 & 2880 & 6917 & 4561 & 5948 & 5319 & 2774 & 3138 & 5599\\
\textbf{\#Edges} & 39131 & 1593930 & 471918 & 1089250 & 767900 & 288369 & 454160 & 940229 & 112147 & 299886 & 1154406 & 549094 & 1737343 & 1791138 & 248773 & 394605 & 1184264\\
\hline
\end{tabular}
\end{adjustbox}
\end{table*}



\section{Evaluation}
\label{sec:eval}
Our methods are implemented in Python 3.7.9 utilizing the Modin multiprocess library \cite{Petersohn20} and our experiments are conducted in a Windows 10 machine with Intel Core i5-2430, 2.40GHz processor and 8GB RAM. We use three real-world datasets, a collaboration network, \textit{DBLP}, a movie ratings dataset, \textit{MovieLens} \cite{Harper16} and a contact network, \textit{Primary School}. Our code and data are publicly available\footnote{https://github.com/etsoukanara/graphtempo-structure}.

The \textit{DBLP} dataset is extracted from DBLP\footnote{https://dblp.uni-trier.de/}. Each node corresponds to an author and two authors are connected with an edge if they co-author one or more papers in a  year. Our \textit{DBLP} dataset covers a period of 21 years from 2000 to 2020 and we limit it to publications at 21 conferences related to data management. The derived graph is directed and the direction of the edges indicates the order of a paper's authors. 
Each node is associated with one static attribute that is \textit{gender} ($G$) and one time-varying attribute the number of an author's \textit{publications} ($P$) each year. 

The \textit{MovieLens} dataset is built on the benchmark MovieLens \cite{Harper16} movie ratings dataset. We select a period of six months, from May 2000 up to October 2000, where each month corresponds to a time point. 
Each node corresponds to a user and an edge between two users denotes that they have rated the same movie. Our dataset is directed based on the precedence of ratings, and does not contain multiple edges in the unit of time. 
Each node has 3 static attributes: \textit{gender} ($G$), \textit{age} ($A$) with 6 discrete values according to the age group of a user, and \textit{occupation} ($O$) with 21 discrete values, and 1 time-varying attribute: the \textit{average rating} ($R$) of the user per month. 

The \textit{Primary School} \cite{Stehlé11}, \cite{Gemmetto14} dataset is a contact network of 232 students and 10 teachers of a primary school in Lyon, France, describing the face-to-face proximity of students and teachers and covering a period of 17 hours. Each edge denotes a 20-seconds contact, and each node in the network has two static attributes, gender with 3 values: female (F), male (M), and unspecified (U), and class with 11 values, i.e., the school has 5 grades, 1 to 5, with 2 classes each (i.e., 1A, 1B, 2A, 2B, etc) plus teachers.

Tables \ref{tab:dblp}, \ref{tab:movielens}, \ref{tab:pm} show nodes and edges per time point for the \textit{DBLP}, \textit{MovieLens} and \textit{Primary School} dataset respectively. 

\begin{figure*}
\centering
\centering
\begin{subfigure}{0.5\textwidth}
\begin{adjustbox}{width=\textwidth}
\begin{tabular}{c|ccccccccccccccccccccc}
\hline
\centering
\textbf{\#TP} & 2000 & 2001 & 2002 & 2003 & 2004 & 2005 & 2006\\ \hline
\textbf{\#Nodes} & 1707 & 2565 & 1713 & 3602 & 4079 & 7550 & 6100\\
\textbf{\#Edges} & 28180 & 135721 & 23978 & 146819 & 162685 & 481634 & 171002\\ \hline \hline
\textbf{\#TP} & 2007 & 2008 & 2009 & 2010 & 2011 & 2012 & 2013\\ \hline
\textbf{\#Nodes} & 5169 & 6370 & 7787 & 9336 & 7896 & 16064 & 16285\\
\textbf{\#Edges} & 68807 & 92935 & 260132 & 315183 & 191535 & 3255547 & 3106984\\ \hline
\textbf{\#TP} & 2014 & 2015 & 2016 & 2017 & 2018 & 2019 & 2020\\ \hline
\textbf{\#Nodes} & 12310 & 18423 & 22429 & 24720 & 22981 & 55951 & 44889\\
\textbf{\#Edges} & 988132 & 1124155 & 3581241 & 4268275 & 1070694 & 33354625 & 7522430\\ \hline
\end{tabular}
\end{adjustbox}
\caption{Statistics}
\label{fig:dblp-tri.1}
\end{subfigure}\hspace{20mm}%
\begin{subfigure}{0.23\textwidth}
\centering
\includegraphics[width=\textwidth]{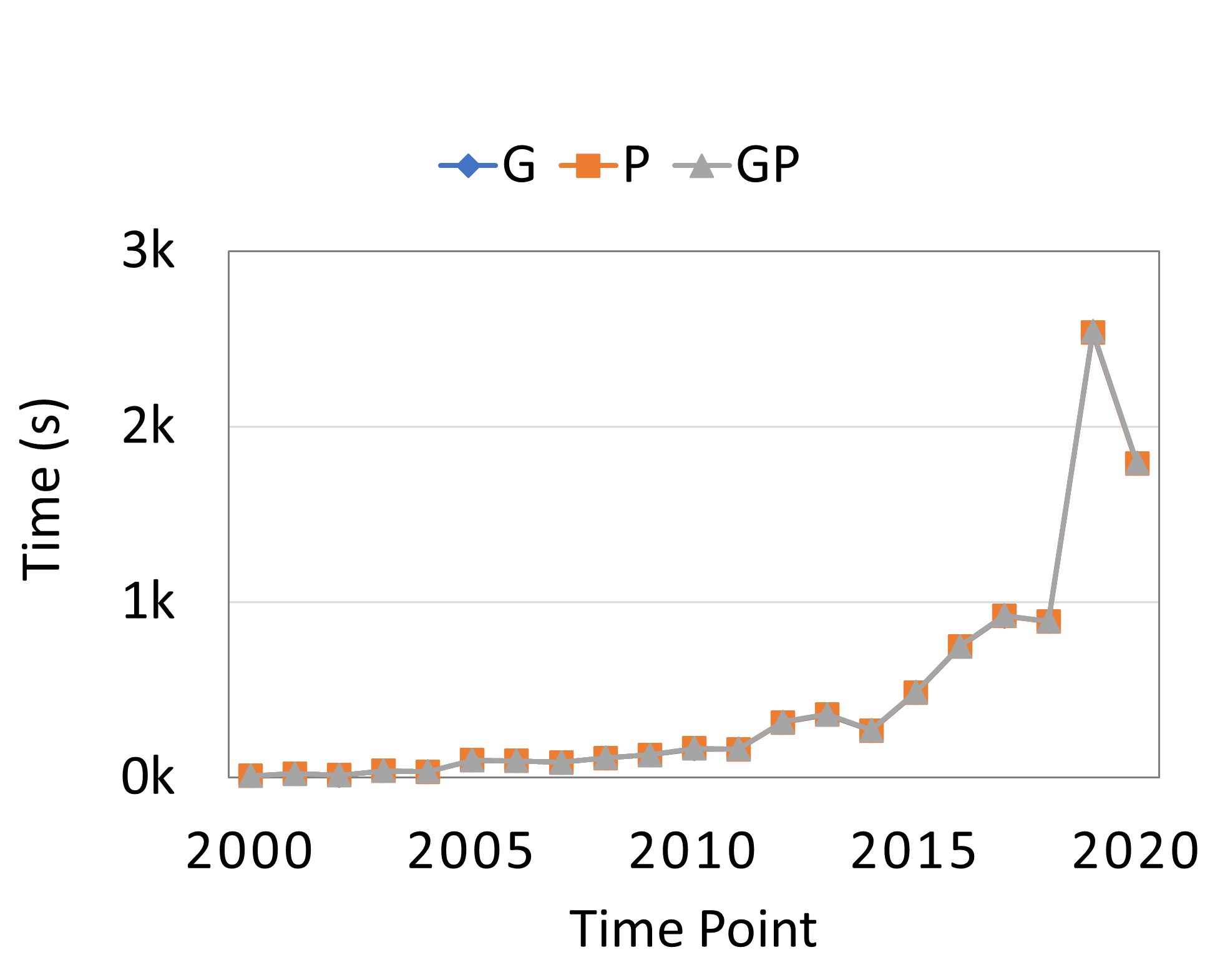}
\caption{Triangles creation time}
\label{fig:dblp-tri.2}
\end{subfigure}
\caption{\textit{DBLP-tri} Graph.}
\label{fig:dblp-tri}
\end{figure*}

With regards to pattern aggregation, we study triangle based aggregation. In Fig. \ref{fig:dblp-tri.1}, we show the number of triangles, that is the nodes, and edges for each time point for the DBLP dataset, while Fig. \ref{fig:dblp-tri.2} reports construction time for the corresponding tri-graph. 
We can see that construction is time-consuming and the tri-graph size is much greater than the original graphs. Table  \ref{tab:pmtri} reports the corresponding nodes and edges for the Primary School tri-graph.

\begin{figure*}
\centering
\begin{subfigure}{0.23\textwidth}
\includegraphics[width=\textwidth]{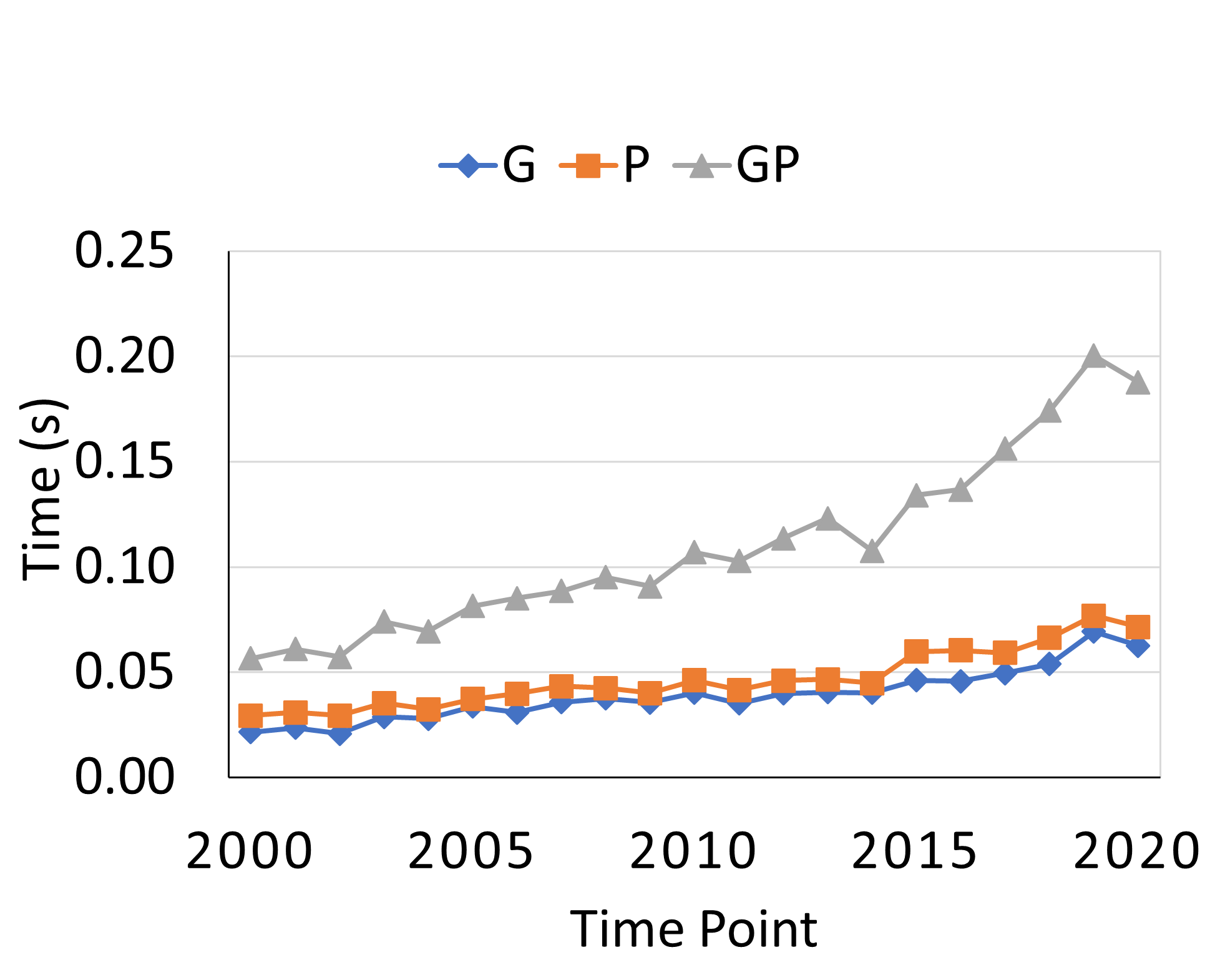}
\caption{\textit{DBLP}}
\label{fig:tps.1}
\end{subfigure}
\begin{subfigure}{0.23\textwidth}
\includegraphics[width=\textwidth]{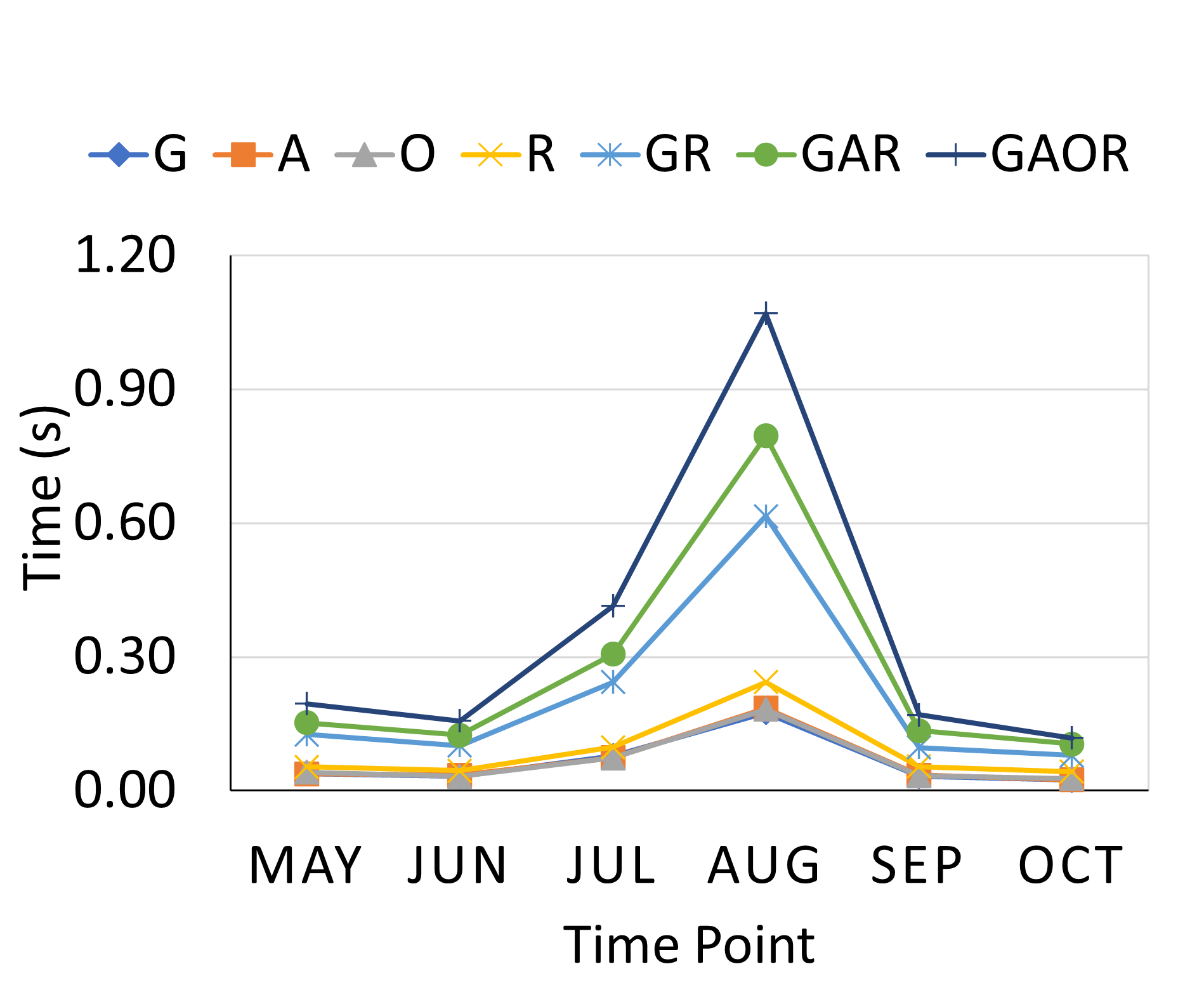}
\caption{\textit{MovieLens}}
\label{fig:tps.2}
\end{subfigure}
\begin{subfigure}{0.23\textwidth}
\includegraphics[width=\textwidth]{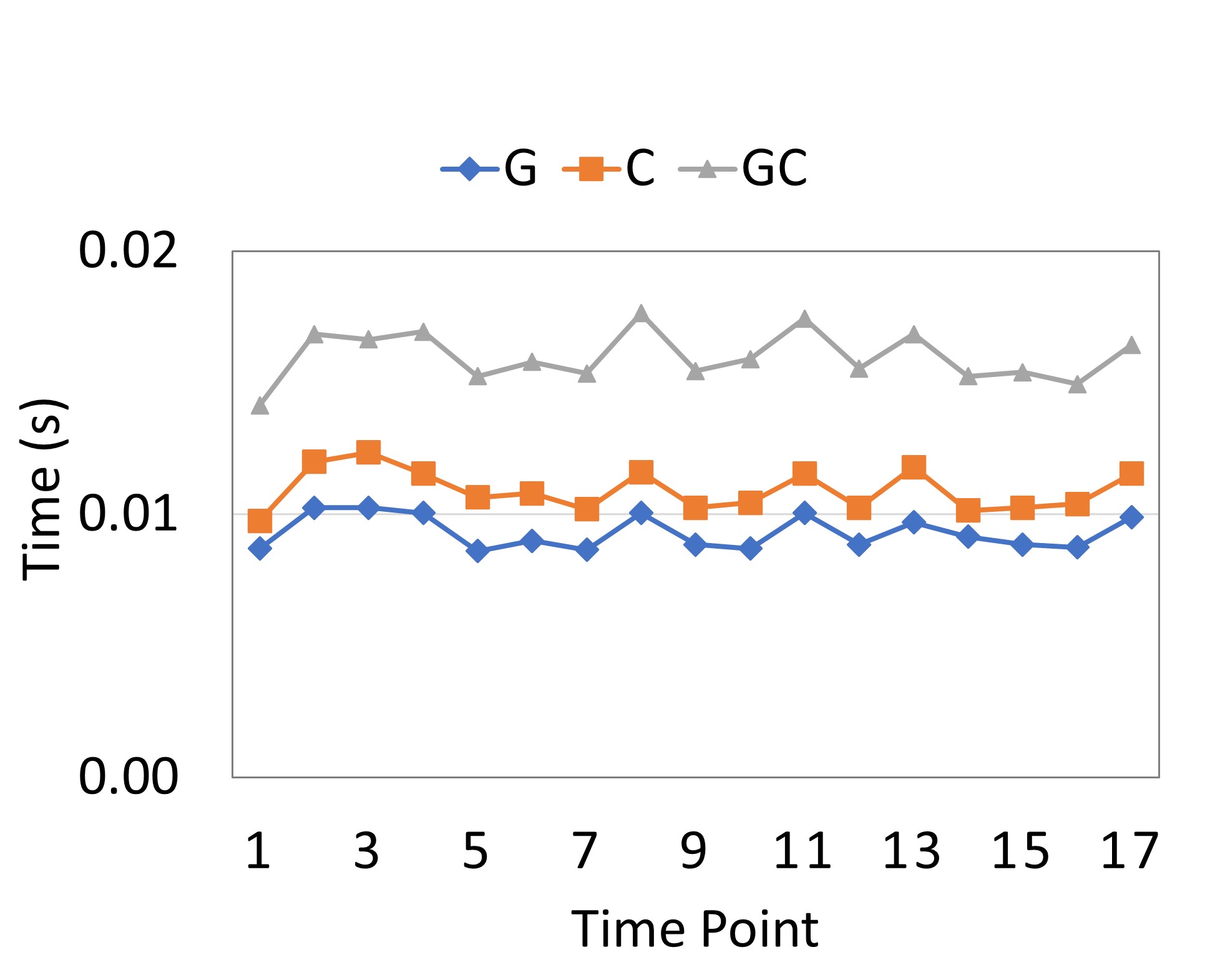}
\caption{\textit{Primary School}}
\label{fig:tps.3}
\end{subfigure}
\begin{subfigure}{0.23\textwidth}
\includegraphics[width=\textwidth]{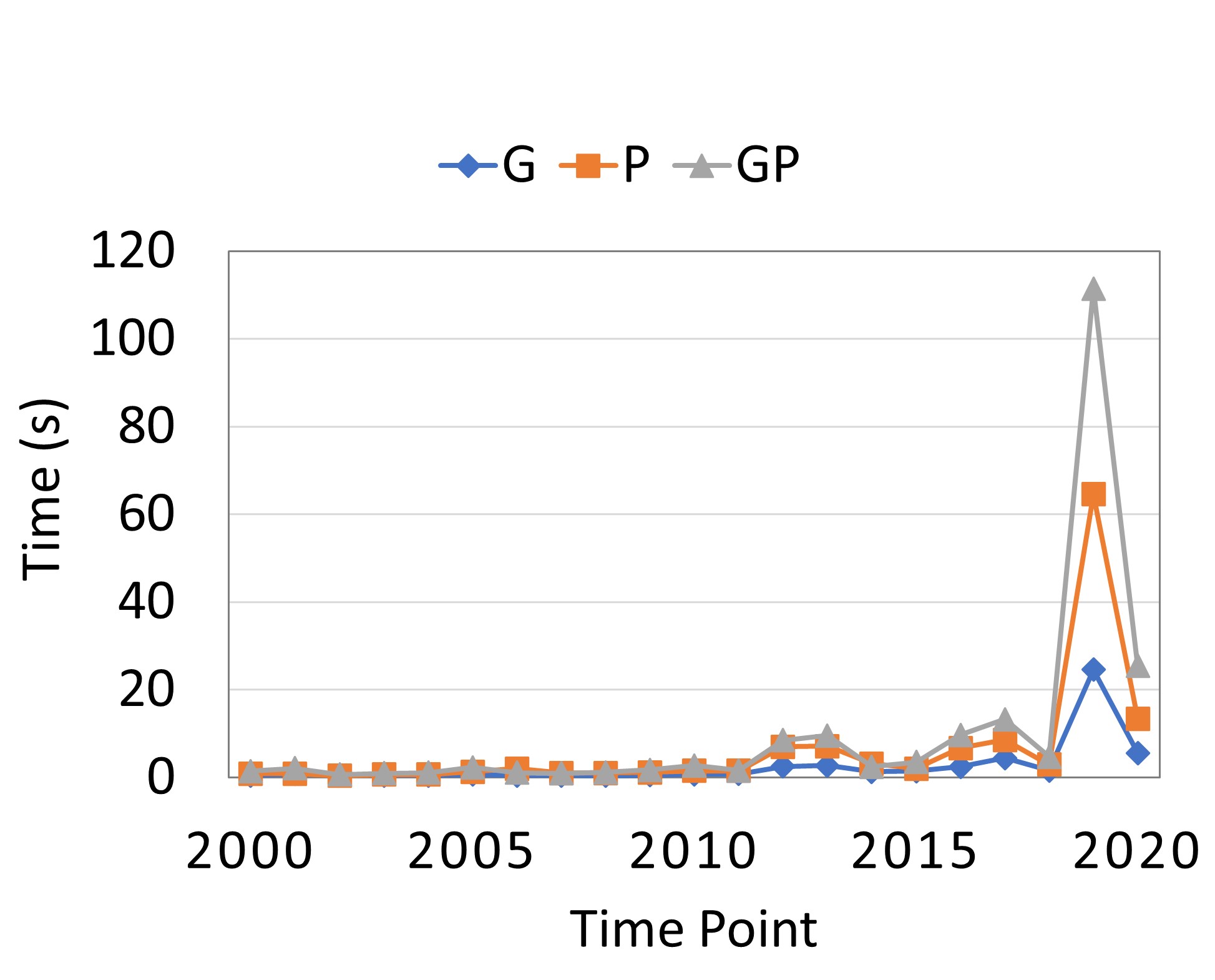}
\caption{\textit{DBLP-tri}}
\label{fig:tps.4}
\end{subfigure}
\caption{Aggregation time per attribute and groups of attributes per time point.}
\label{fig:tps}
\end{figure*}

\begin{figure*}
\centering
\begin{subfigure}{0.23\textwidth}
\includegraphics[width=\textwidth]{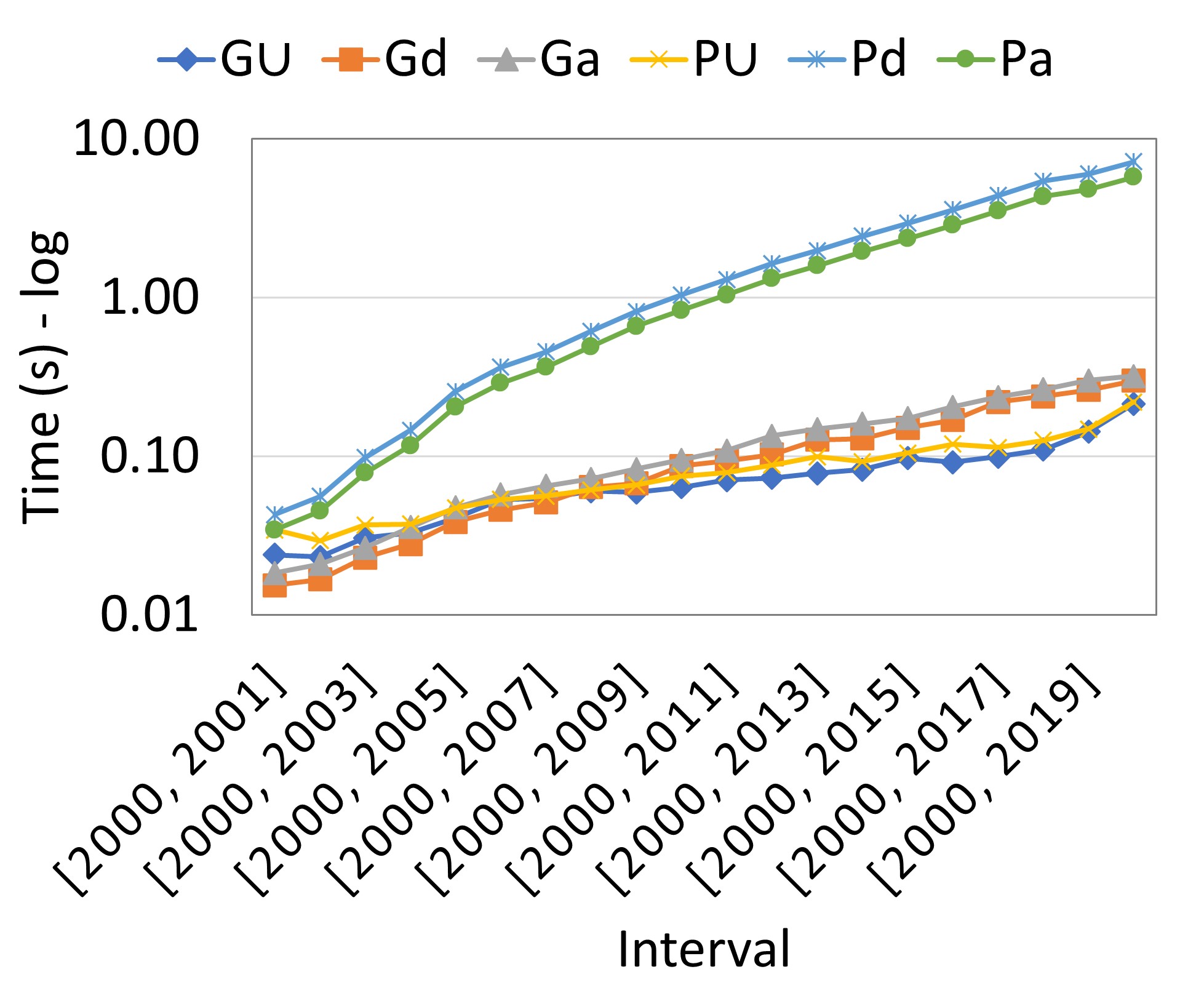}
\caption{Union}
\label{fig:operdblp.1}
\end{subfigure}
\begin{subfigure}{0.23\textwidth}
\includegraphics[width=\textwidth]{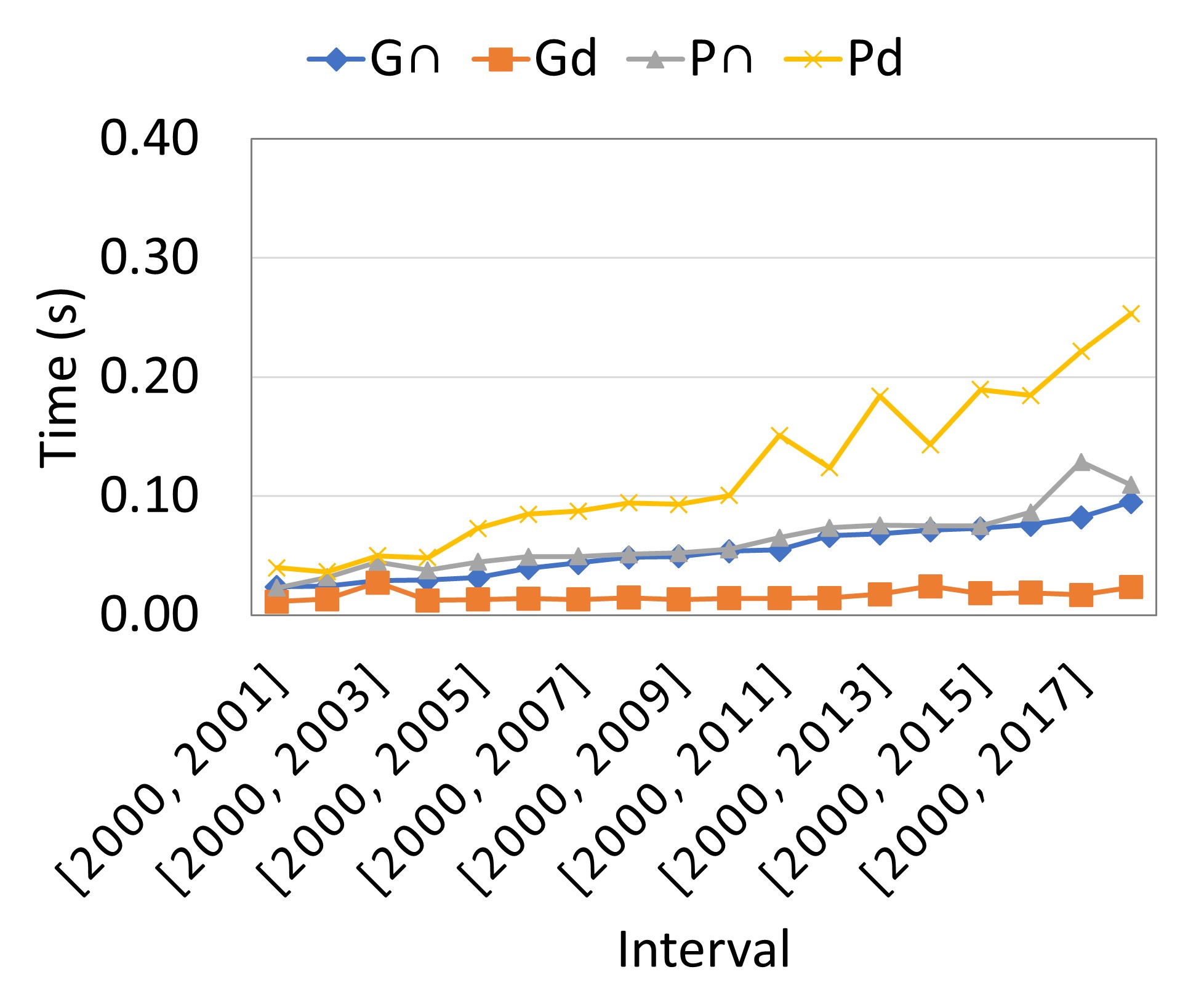}
\caption{Intersection}
\label{fig:operdblp.2}
\end{subfigure}
\begin{subfigure}{0.23\textwidth}
\includegraphics[width=\textwidth]{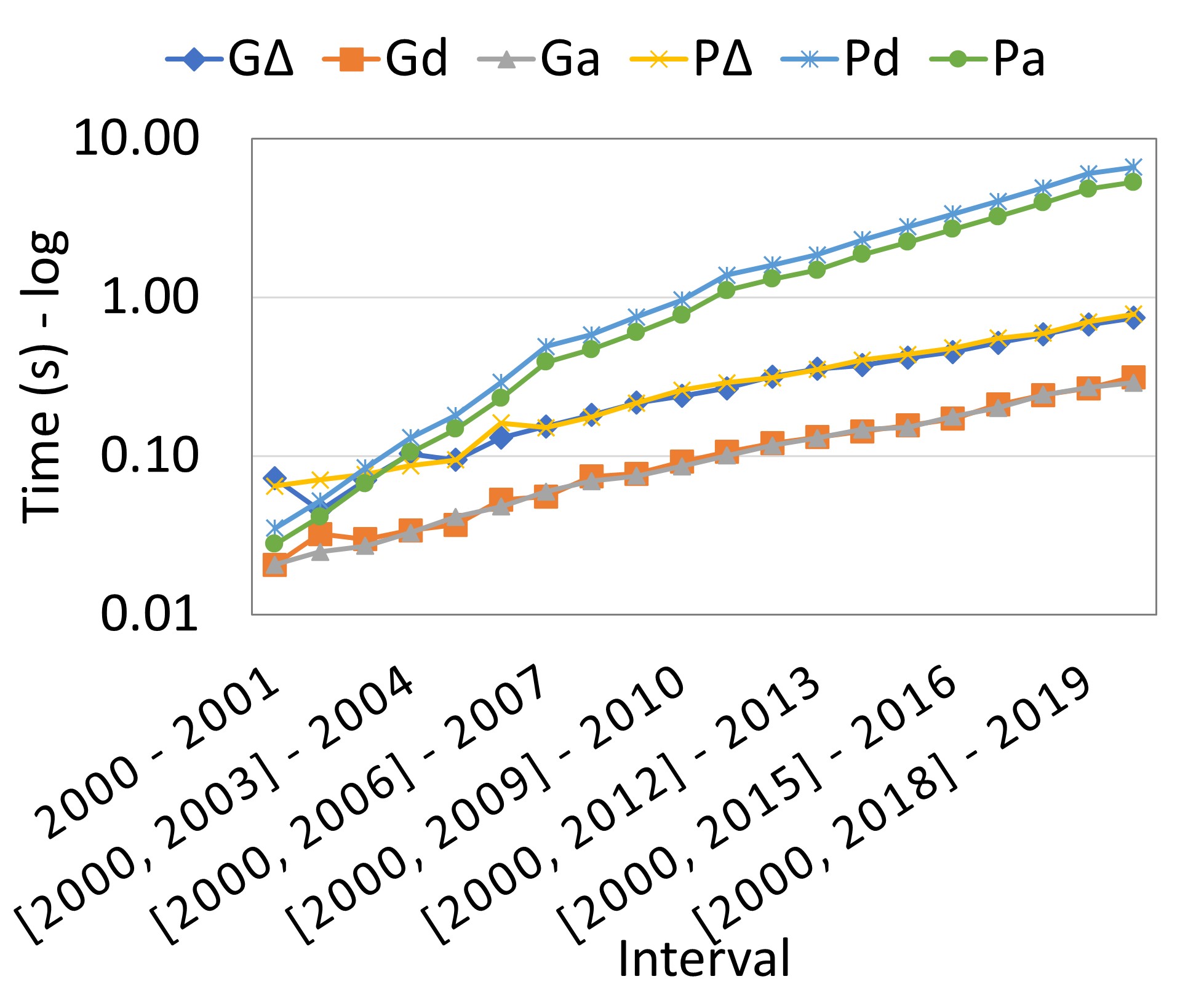}
\caption{Difference ($\mathcal{T}_{old}$($\cup$) - $\mathcal{T}_{new}$)}
\label{fig:operdblp.3}
\end{subfigure}
\begin{subfigure}{0.23\textwidth}
\includegraphics[width=\textwidth]{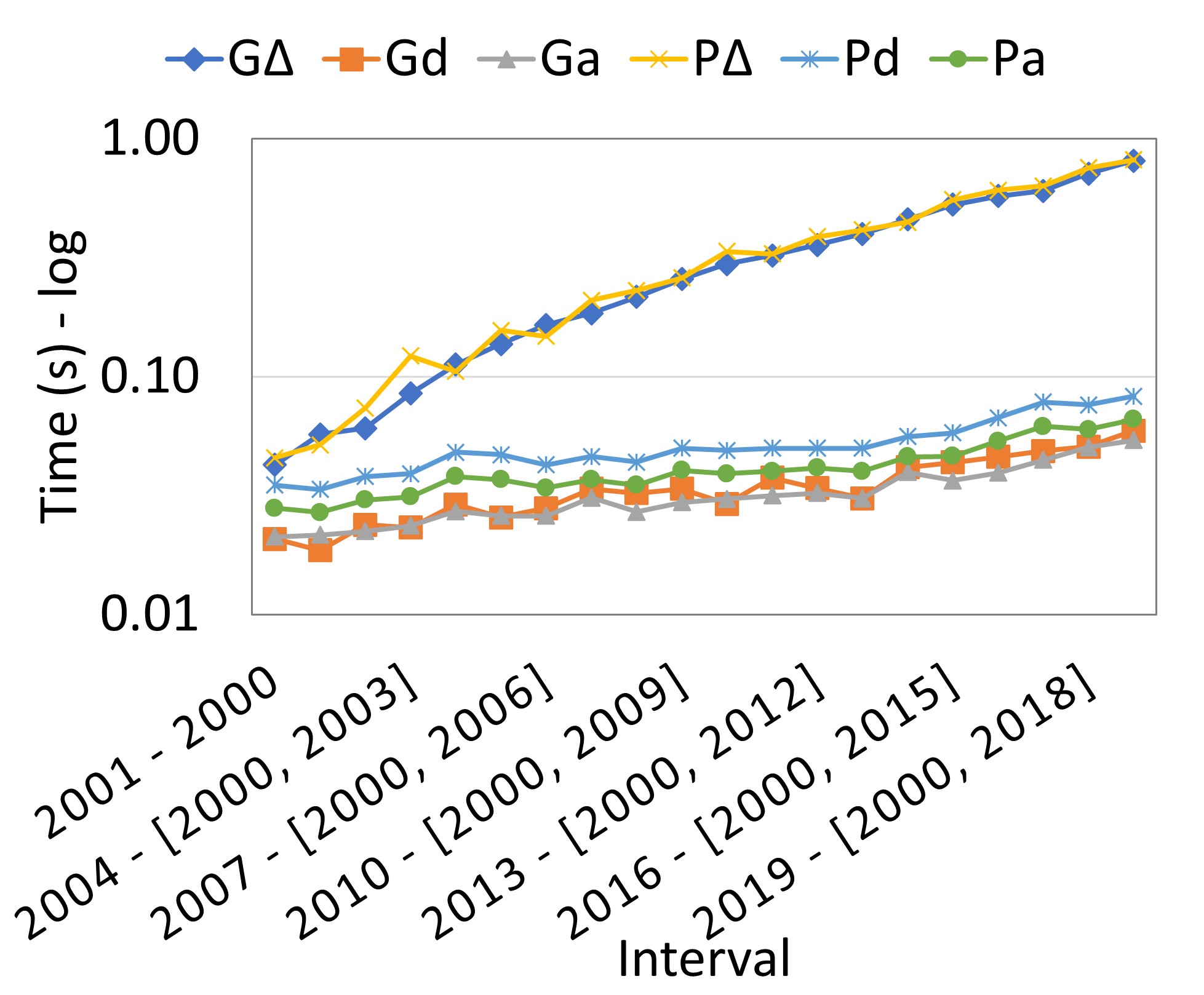}
\caption{Difference ($\mathcal{T}_{new}$ - $\mathcal{T}_{old}$($\cup$))}
\label{fig:operdblp.4}
\end{subfigure}
\caption{Temporal and attribute aggregation for \textit{DBLP}.}
\label{fig:operdblp}
\end{figure*}

\begin{figure}
\centering
\begin{subfigure}{0.23\textwidth}
\includegraphics[width=\textwidth]{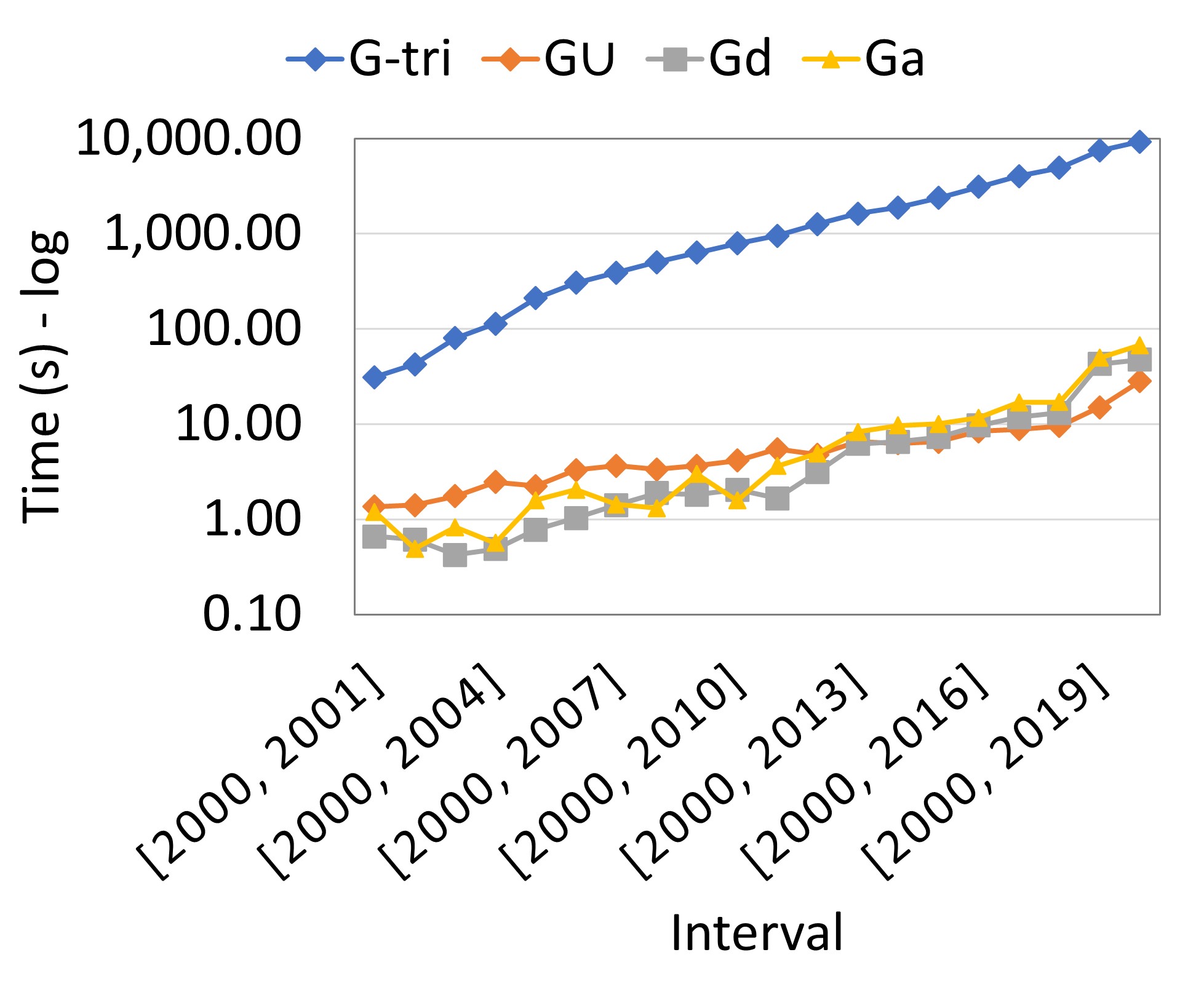}
\caption{Union}
\label{fig:opertri.1}
\end{subfigure}
\begin{subfigure}{0.23\textwidth}
\includegraphics[width=\textwidth]{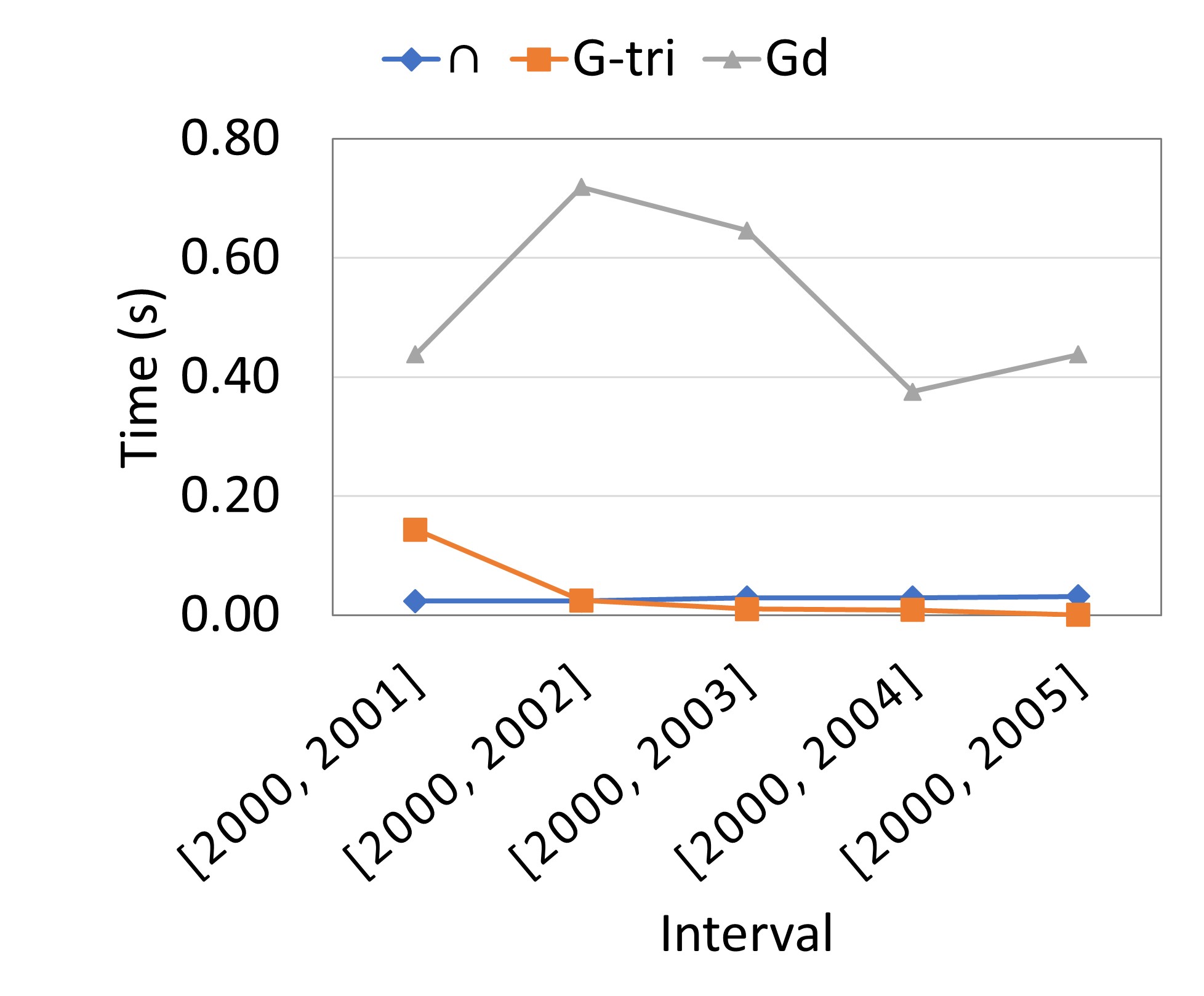}
\caption{Intersection}
\label{fig:opertri.2}
\end{subfigure}
\caption{Temporal and pattern aggregation for \textit{DBLP-tri}.}
\label{fig:opertri}
\end{figure}

\begin{figure*}
\centering
\begin{subfigure}{0.23\textwidth}
\includegraphics[width=\textwidth]{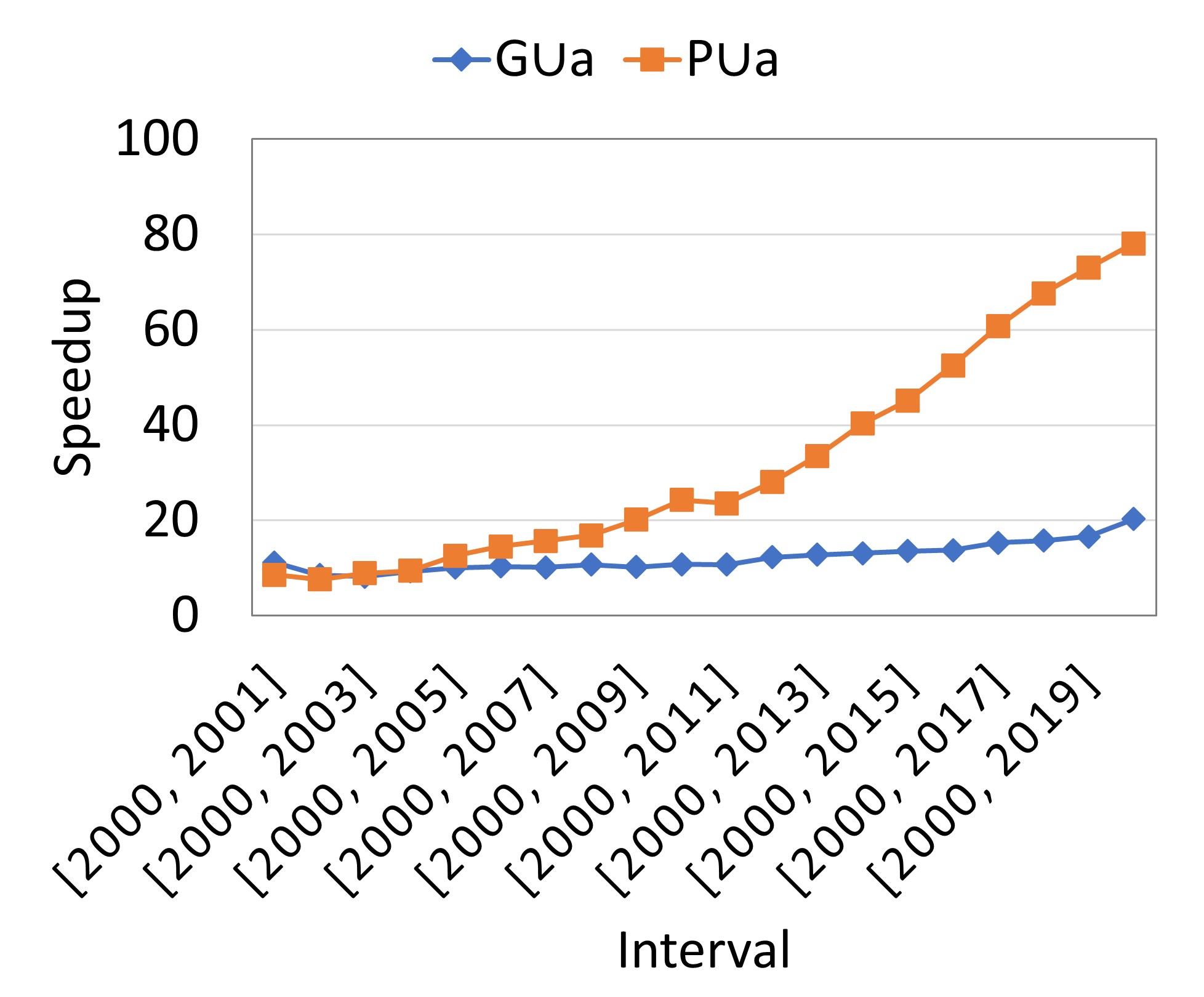}
\caption{\textit{DBLP}}
\label{fig:spupu.1}
\end{subfigure}\hspace{10mm}
\begin{subfigure}{0.23\textwidth}
\includegraphics[width=\textwidth]{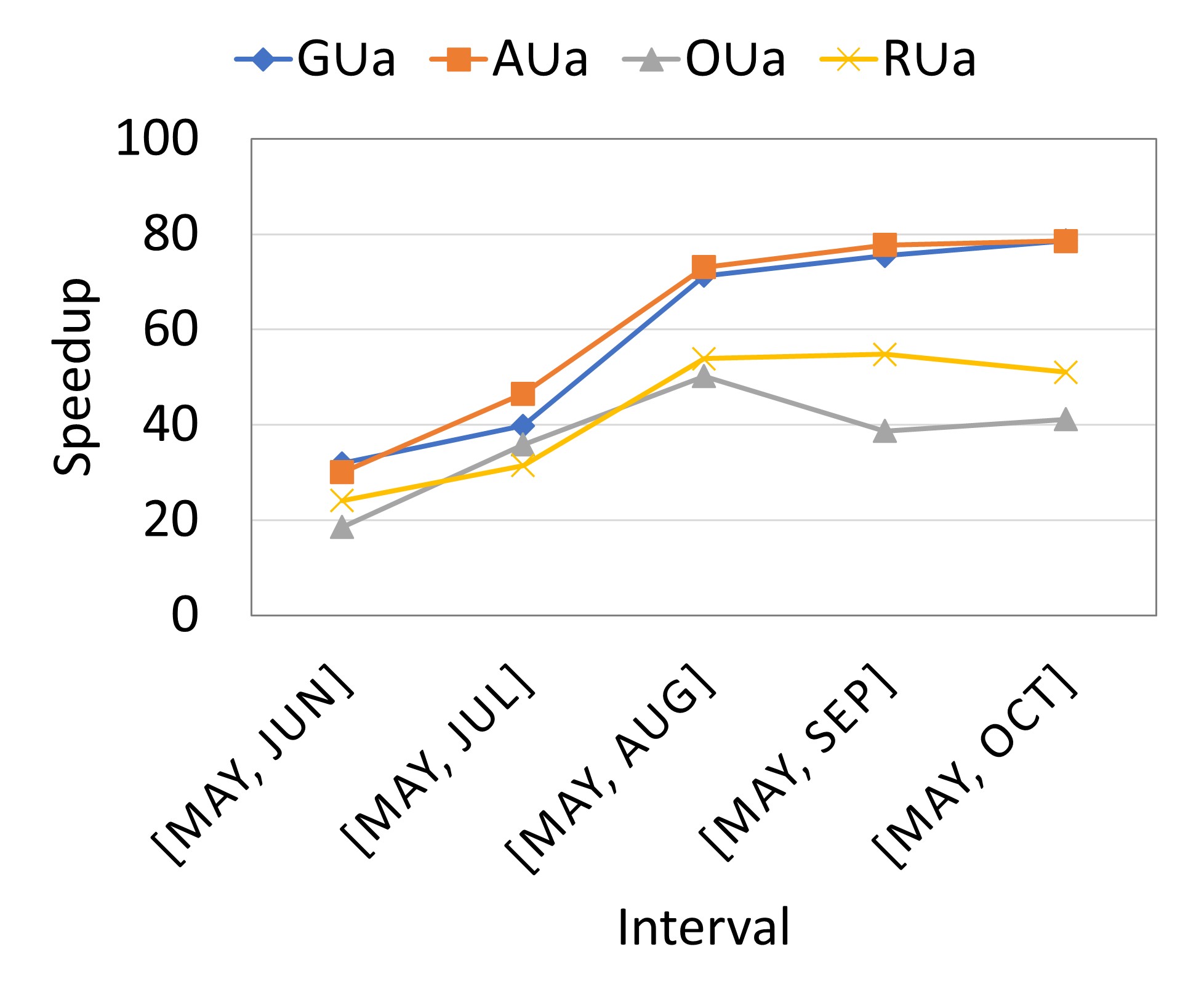}
\caption{\textit{MovieLens}}
\label{fig:spupu.2}
\end{subfigure}\hspace{15mm}
\begin{subfigure}{0.23\textwidth}
\includegraphics[width=\textwidth]{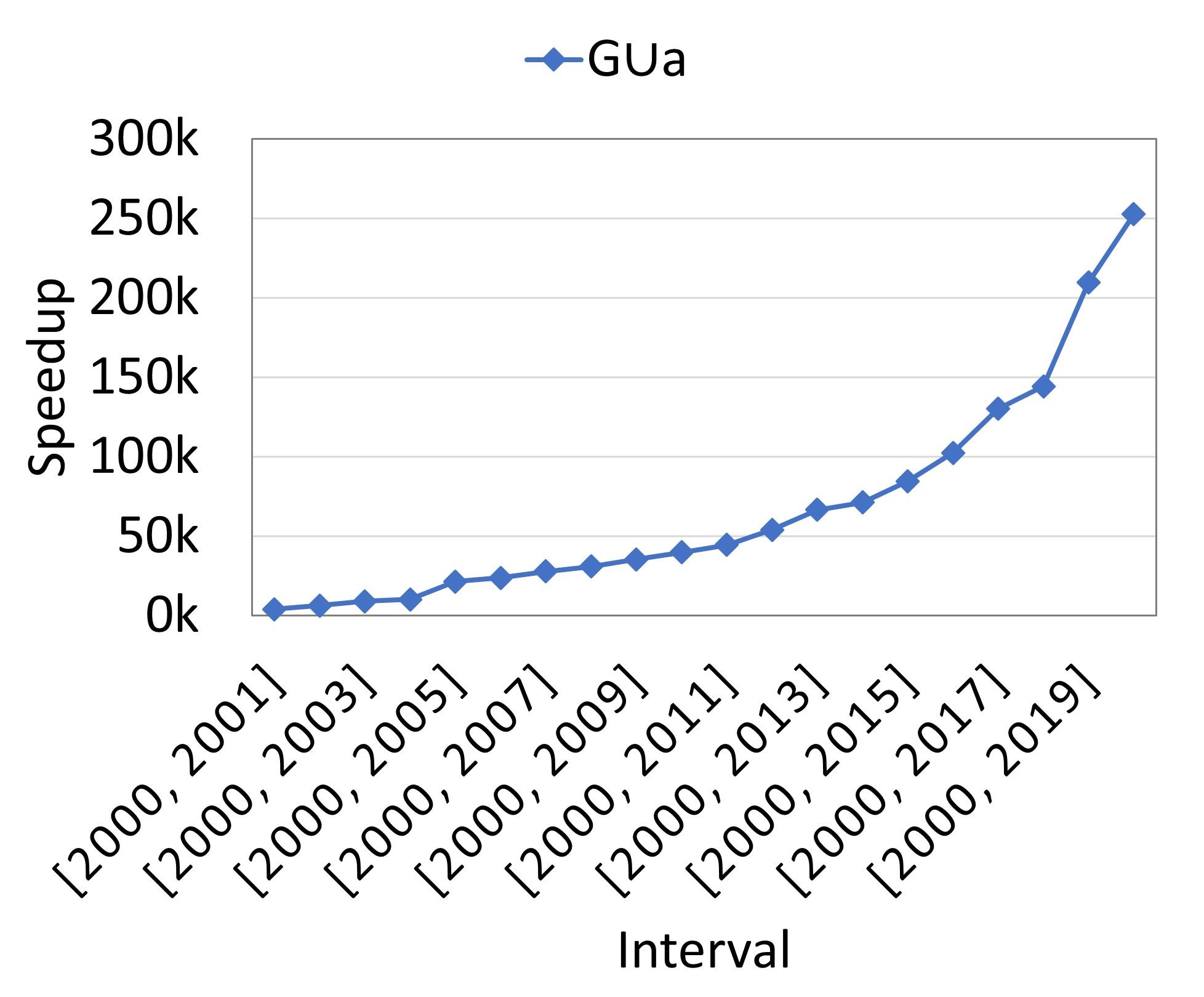}
\caption{\textit{DBLP-tri}}
\label{fig:spupu.3}
\end{subfigure}
\caption{Speedup of efficient Union aggregation (ALL).}
\label{fig:spupu}
\end{figure*}

\begin{figure}
\centering
\begin{subfigure}{0.24\textwidth}
\includegraphics[width=\textwidth]{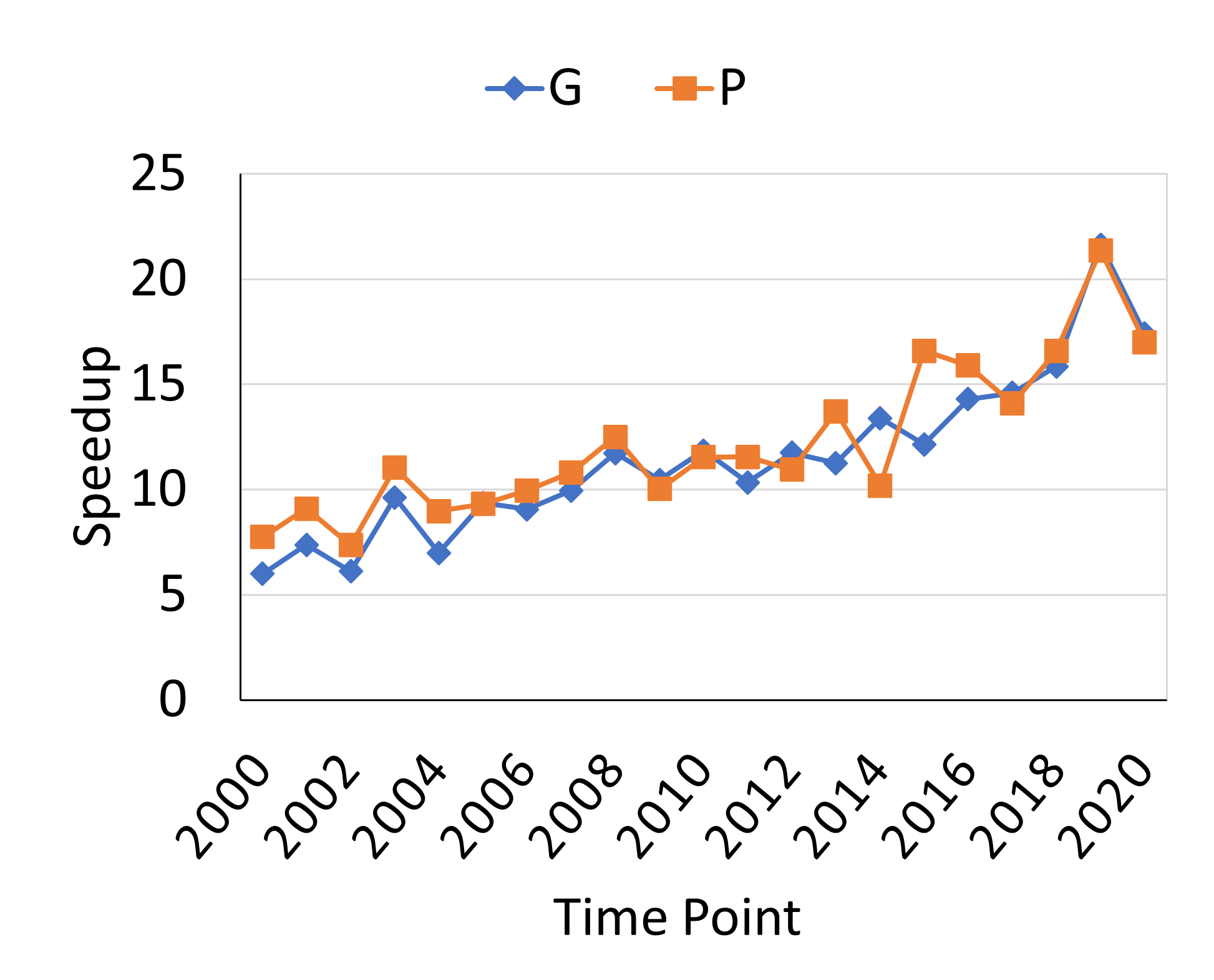}
\caption{Single attributes}
\label{fig:spupdim.1}
\end{subfigure}
\begin{subfigure}{0.24\textwidth}
\includegraphics[width=\textwidth]{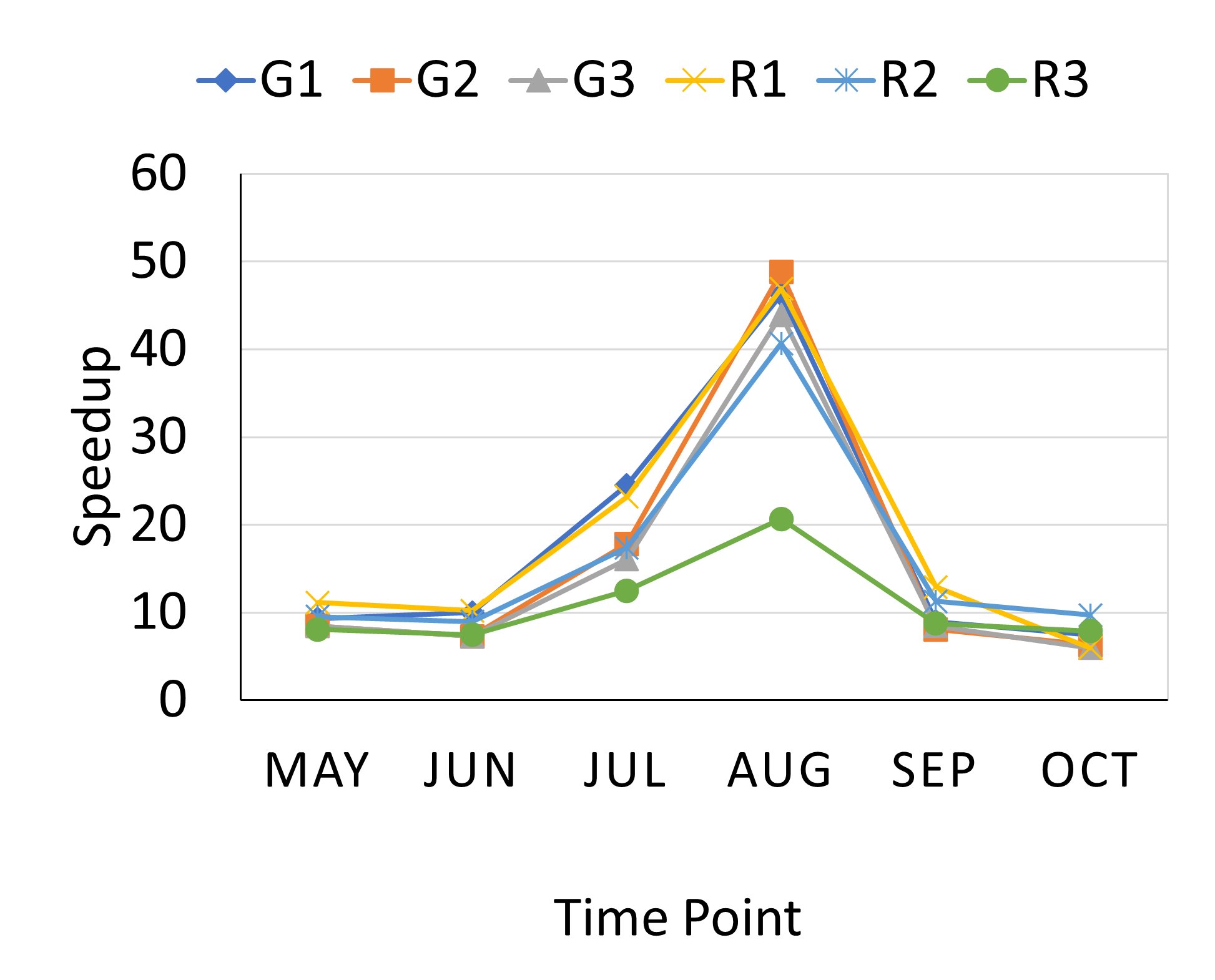}
\caption{Single attributes}
\label{fig:spupdim.2}
\end{subfigure}
\begin{subfigure}{0.24\textwidth}
\includegraphics[width=\textwidth]{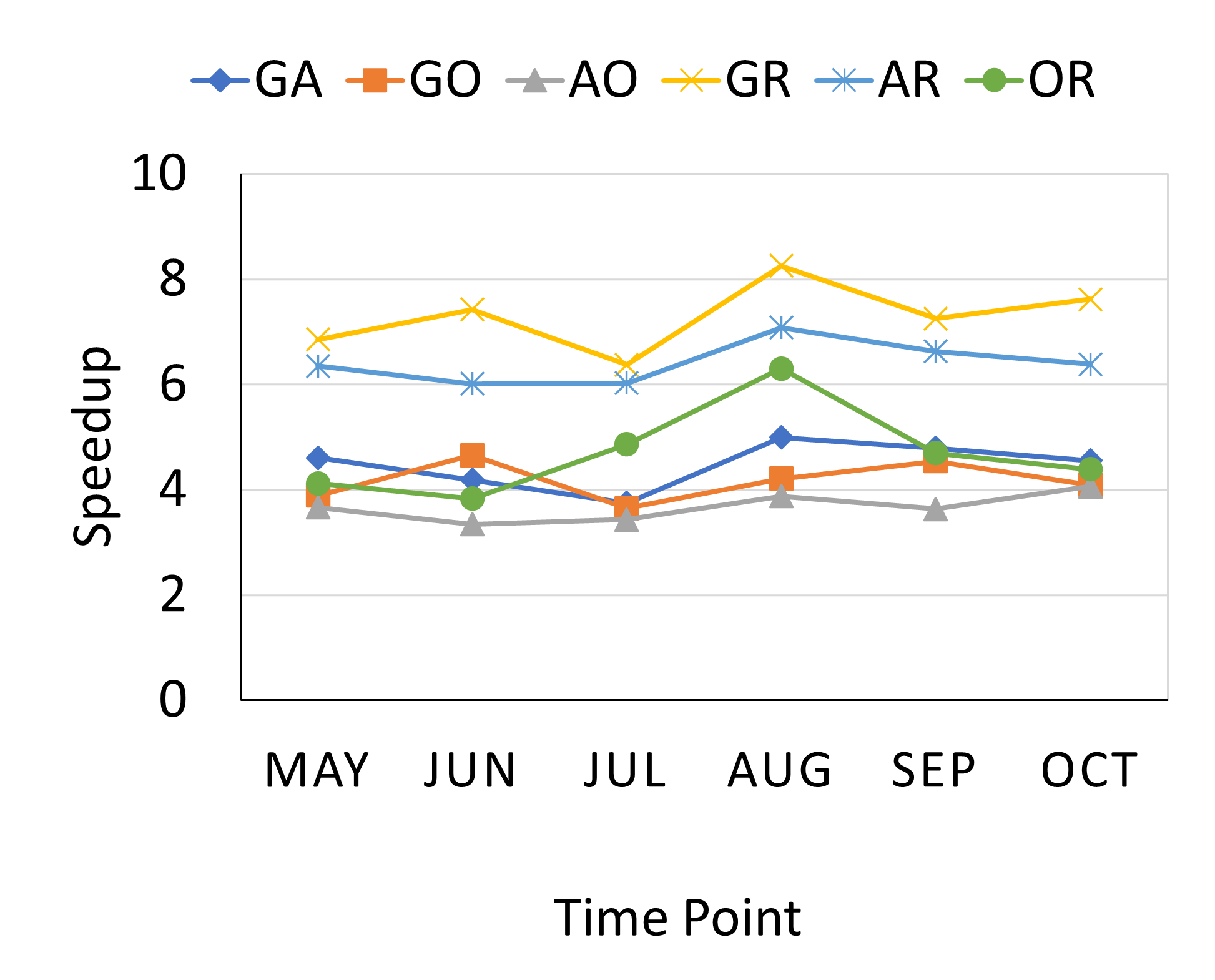}
\caption{Pairs of attributes}
\label{fig:spupdim.3}
\end{subfigure}
\begin{subfigure}{0.24\textwidth}
\includegraphics[width=\textwidth]{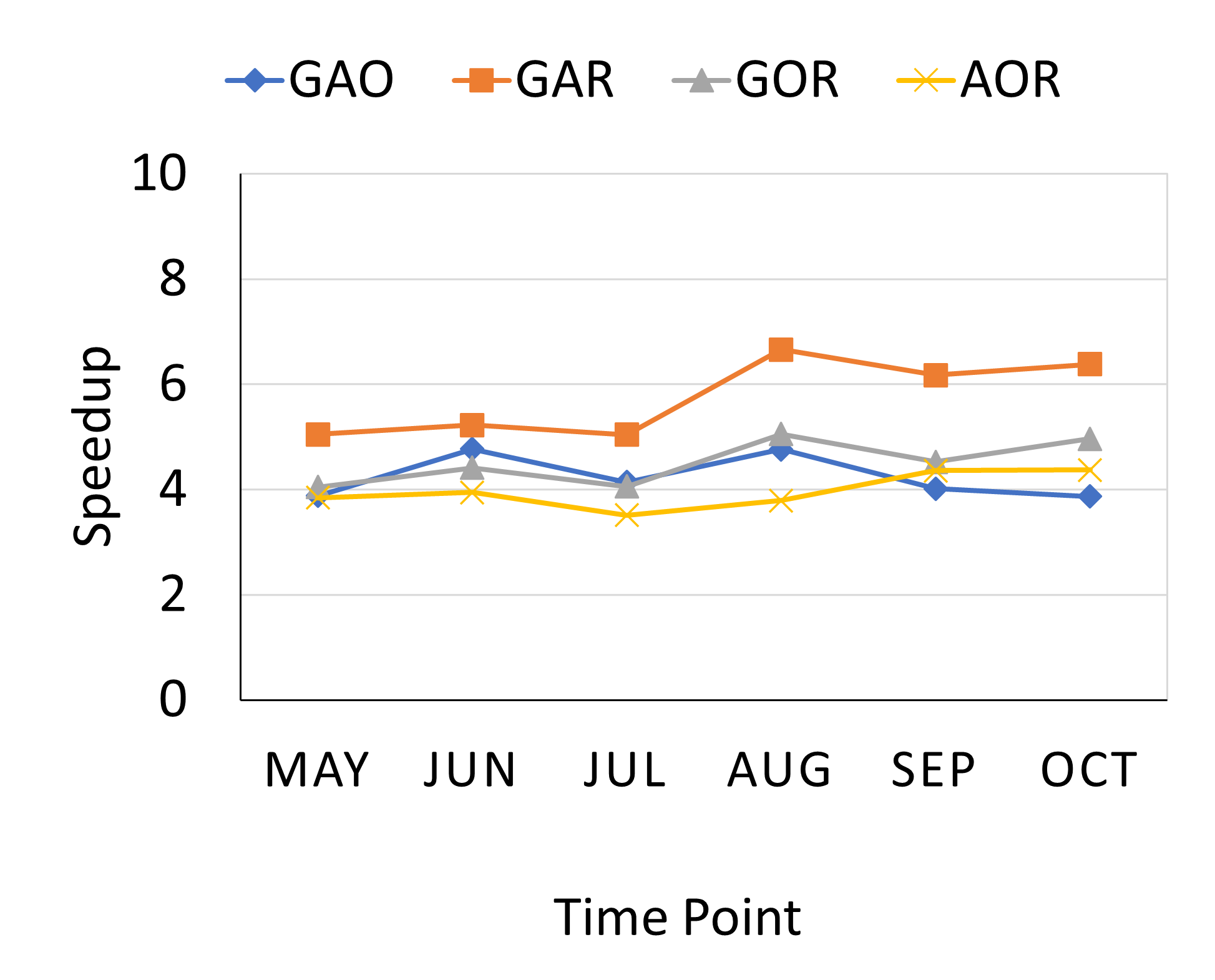}
\caption{Triples of attributes}
\label{fig:spupdim.4}
\end{subfigure}
\begin{subfigure}{0.24\textwidth}
\includegraphics[width=\textwidth]{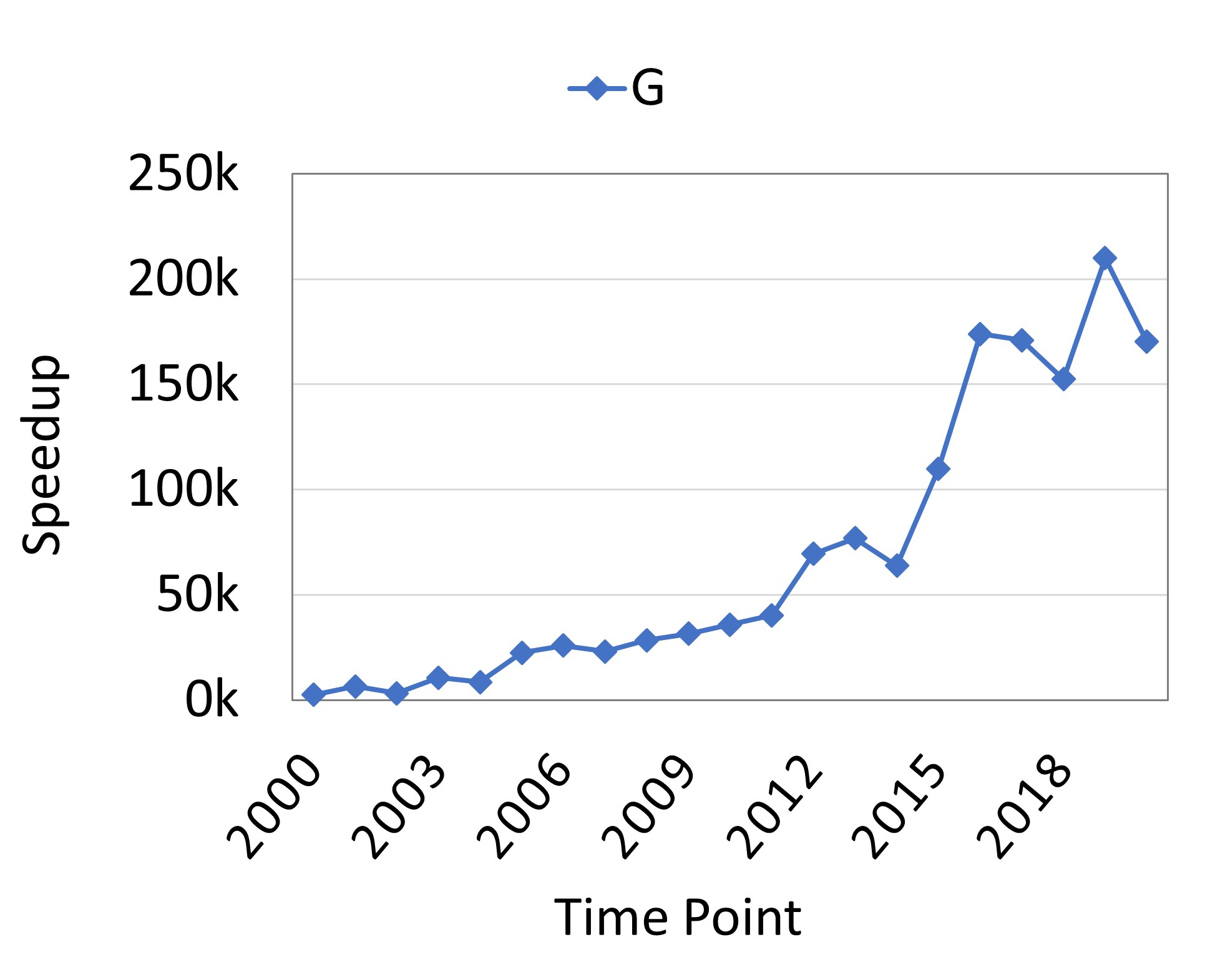}
\caption{Single attributes}
\label{fig:spupdim.5}
\end{subfigure}
\caption{Speedup of efficient attribute aggregation per time point for (a) \textit{DBLP}, (b-d) \textit{MovieLens}, (e) \textit{DBLP-tri}.}
\label{fig:spupdim}
\end{figure}

\begin{figure}
\centering
\begin{subfigure}{0.45\textwidth}
\centering
\includegraphics[width=1\textwidth]{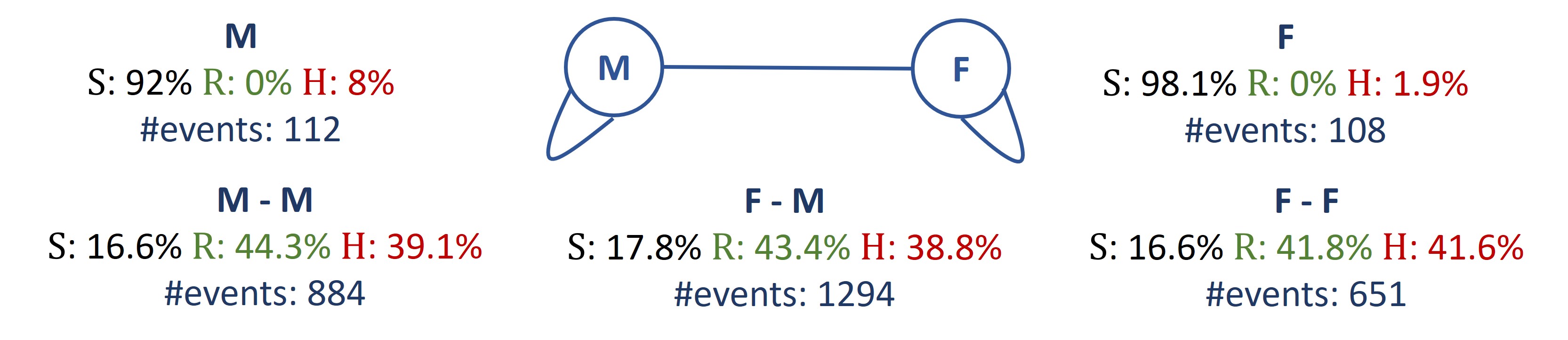}
\caption{\textit{Primary School}}
\label{fig:f14.1}
\end{subfigure}\hspace{10mm}
\begin{subfigure}{0.45\textwidth}
\centering
\includegraphics[width=1\textwidth]{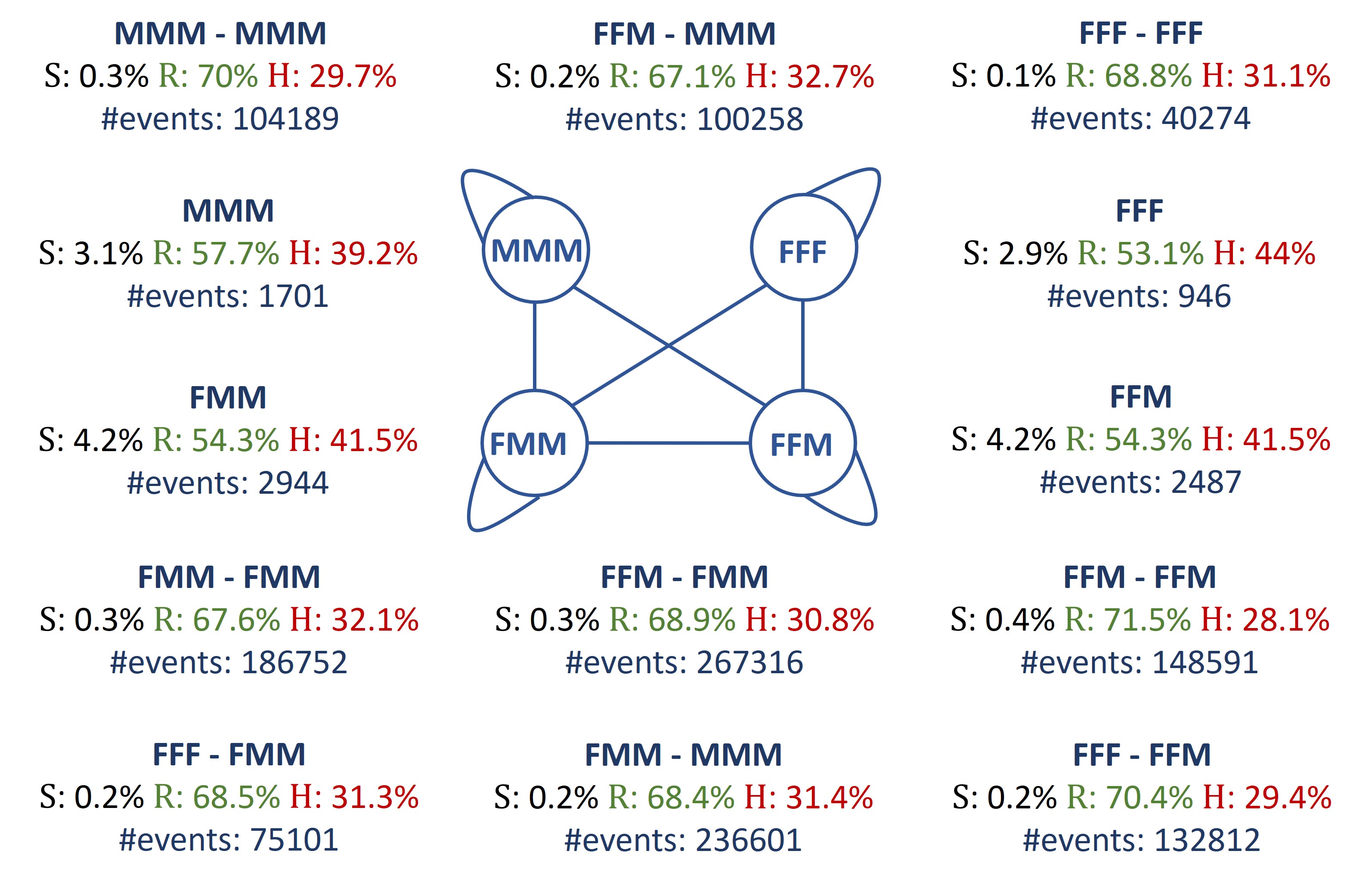}
\caption{\textit{Primary School}-tri}
\label{fig:f14.2}
\end{subfigure}
\caption{Evolution on Gender for the 4th (break) as to 3rd (lesson) time point.}
\label{fig:f14}
\end{figure}



\begin{figure*}
\centering
\begin{subfigure}{0.22\textwidth}
\centering
\includegraphics[width=\textwidth]{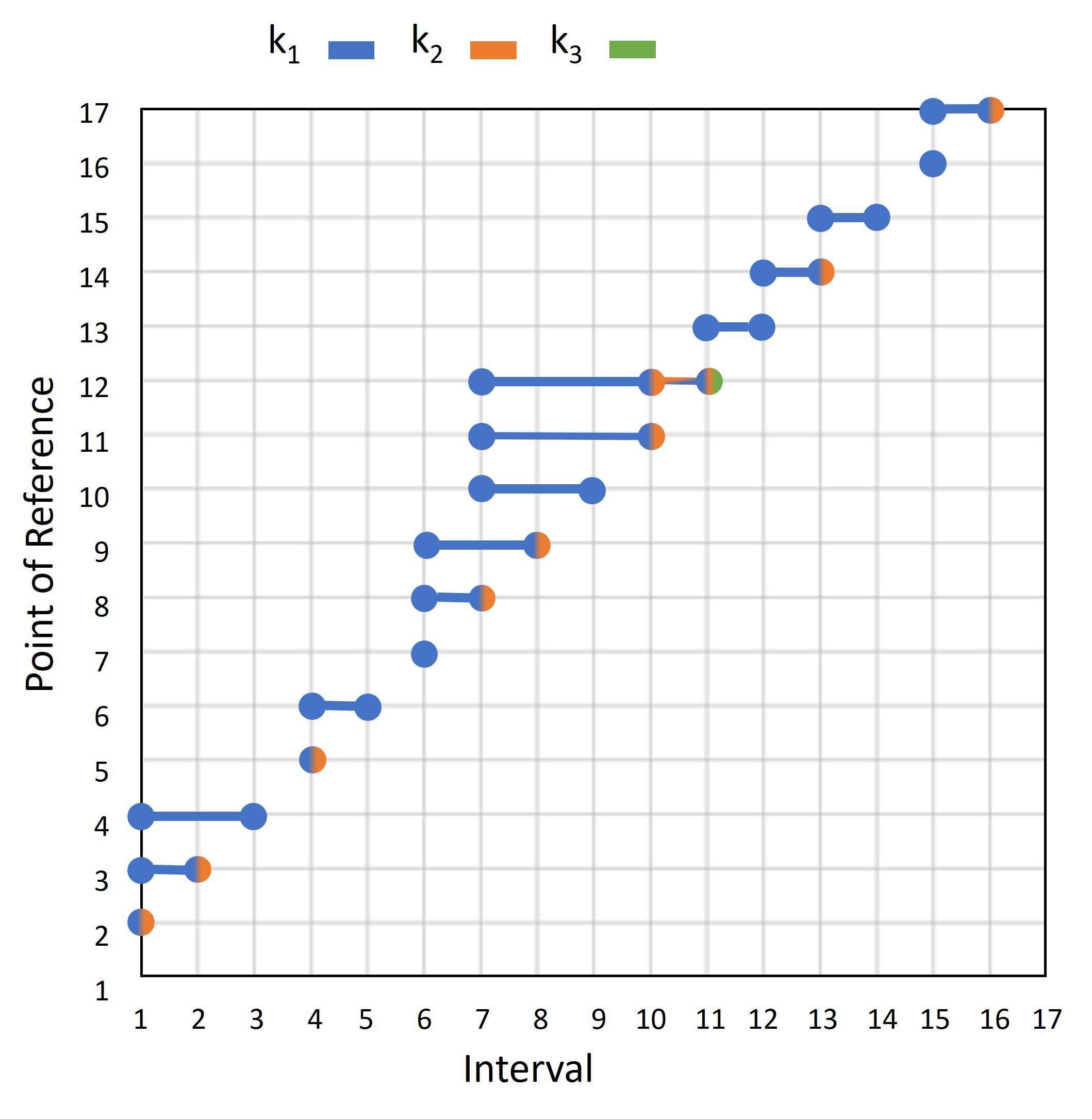}
\caption{Stability ($\mathcal{T}_{old}$($\cap$) $\cap$ $\mathcal{T}_{new}$)}
\label{fig:f15.1}
\end{subfigure}\hspace{15mm}%
\begin{subfigure}{0.22\textwidth}
\centering
\includegraphics[width=\textwidth]{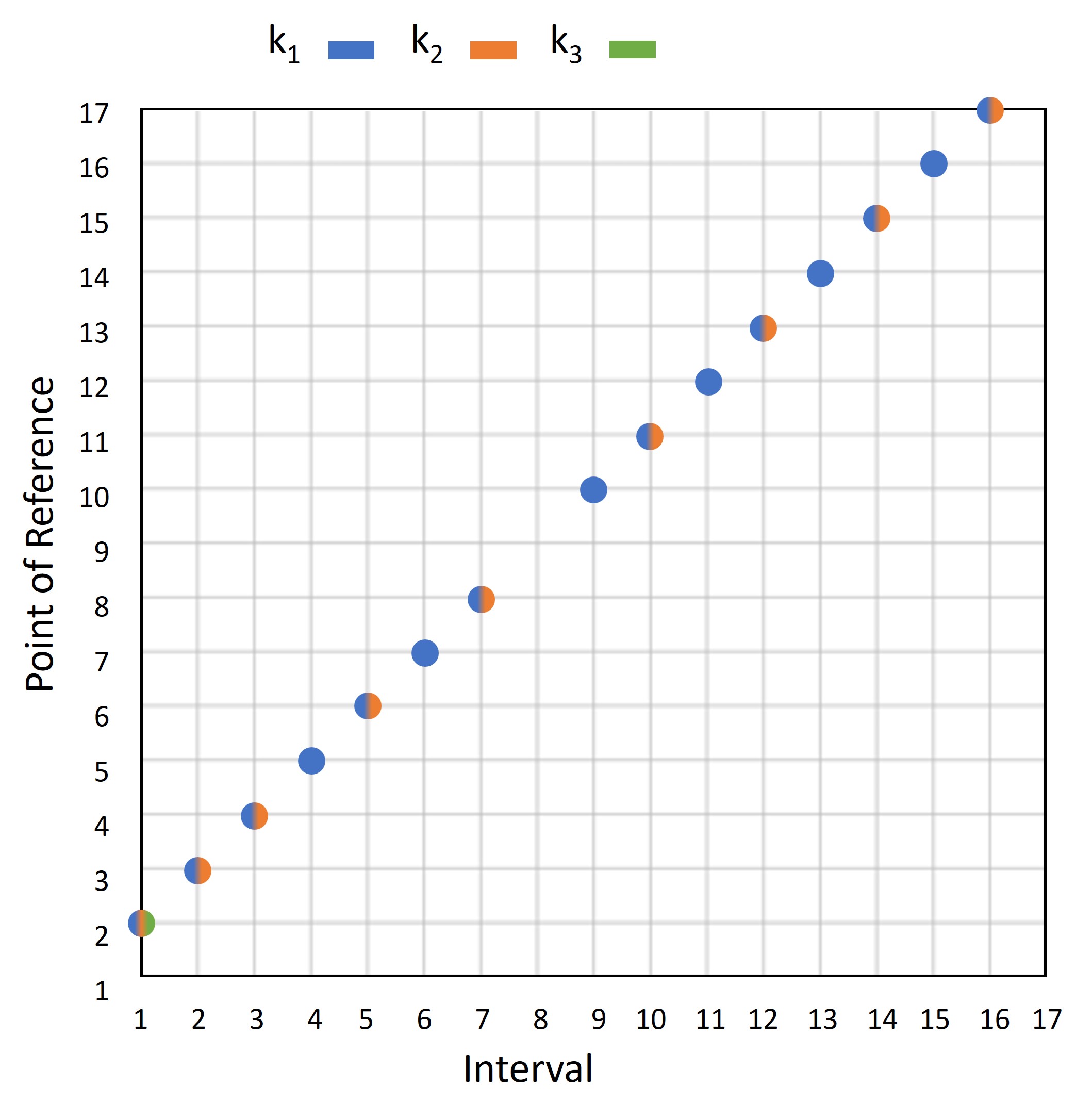}
\caption{Growth ($\mathcal{T}_{new}$ - $\mathcal{T}_{old}$($\cup$))}
\label{fig:f15.2}
\end{subfigure}\hspace{15mm}%
\begin{subfigure}{0.22\textwidth}
\centering
\includegraphics[width=\textwidth]{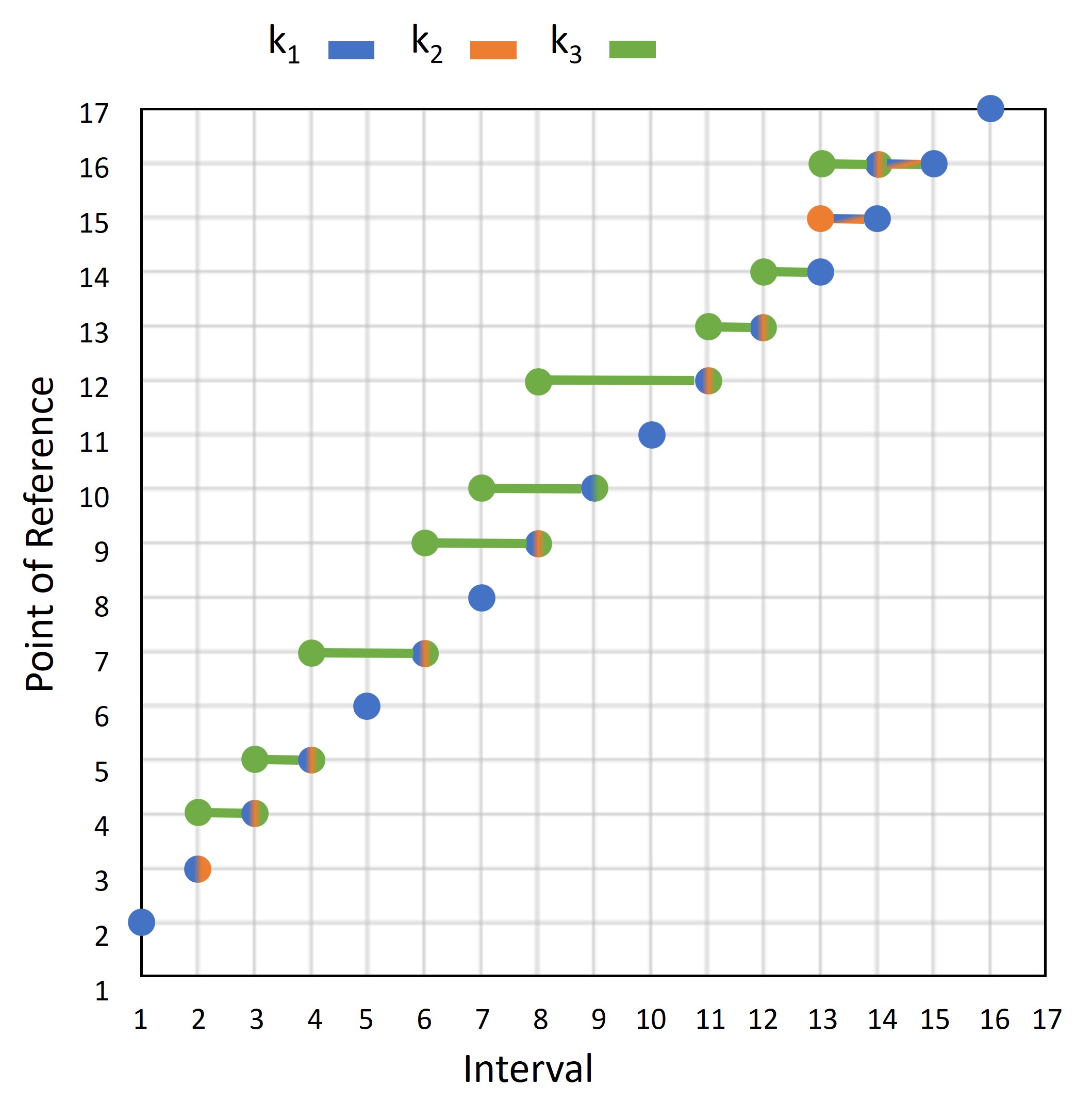}
\caption{Shrinkage ($\mathcal{T}_{old}$($\cup$) - $\mathcal{T}_{new}$)}
\label{fig:f15.3}
\end{subfigure}
\caption{Exploration cases with (a) intersection, and (b), (c) union semantics on $k$ for girls interactions of \textit{Primary School}.}
\label{fig:f15}
\end{figure*}


\subsection{Performance Evaluation}
In the first set of experiments, we evaluate the performance of our algorithms by measuring execution times.

\noindent\textit{Type of Attribute.} 
In Fig. \ref{fig:tps}, we measure aggregation time per attribute and all attributes combinations on time points. While we expect static attributes to be faster compared to time-varying ones, in Fig. \ref{fig:tps.1}, for \textit{DBLP} time is similar for both attributes, since both aggregations have similar behaviour when dealing with time points. Their difference is due to the larger domain of the time-varying attribute, as gender has 2 distinct values, while publications vary from 7 to 18. 
The aggregation (gender, publications) needs 0.18s on 2020, almost 3 times the time for gender (0.06s) and publications (0.07s). When combining attributes, the domain gets larger, e.g. there are 28 distinct values for (gender, publications) forming the number of aggregate nodes. 

In Fig. \ref{fig:tps.2} for \textit{MovieLens}, for presentation clarity, we report part of all possible attribute combinations as the ones omitted behave in a similar way. We depict aggregation time for each attribute, and, from all possible groups with a specific number of attributes, we selectively report a representative example combining static and time-varying attributes. We notice a similar to Fig. \ref{fig:tps.1} pattern as gender aggregation hits the best time overall, while aggregation on all 4 attributes has the longest time. The results confirm that time is analogous to the distinct values in the aggregation attribute or combination of attributes domain. The peak observed during August is due to its high number of nodes and edges compared to other months (Table \ref{tab:movielens}).

For the \textit{Primary School} in Fig. \ref{fig:tps.3}, we see a similar behavior with the aggregation for gender requiring the minimum time, while the aggregation for the combination of attributes is the most time-consuming.
The graph size is well-balanced across time, thus, we see almost no change on performance.

For pattern aggregation, we report time for the DBLP tri-graph (Fig.  \ref{fig:tps.4}). Note that aggregation is performed on the tri-graph whose construction as we saw dominates the time consumption, e.g., it reaches 2541s in 2019 (Fig. \ref{fig:dblp-tri.2}). Aggregation is comparatively much faster, only reaching up to 111s for the attribute combination in 2019, but is still slower compared to the other cases as the tri-graph size is also much larger.


\noindent\textit{Temporal Operators.} Figures \ref{fig:operdblp} and \ref{fig:opertri} focus on temporal aggregations for different operators comparing static and time-varying attributes while extending time intervals. We report the time of operation and aggregation separately per attribute for \textit{DBLP}, and \textit{DBLP-tri}.

Figure \ref{fig:operdblp.1} illustrates results for union ($\cup$) in logarithmic scale and compares non-distinct ($a$) and distinct ($d$) aggregation for \textit{DBLP}. The type of attribute greatly influences behavior. Non-distinct aggregation for the longest interval needs 0.53s for gender, while 5.9s for the time-varying attribute. This difference is observed because for time-varying attributes, we need to capture all their different values in the interval.
Figure \ref{fig:operdblp.2} depicts intersection ($\cap$), where the results refer up to [2000, 2017], indicating the longest interval there exists at least one common edge. The type of attribute is again the controlling performance factor. In contrast to union, intersection is more costly than aggregation for the static attribute. This is due to the decreasing graph size as the intervals expand. For time-varying attributes, the operation time is similar to static, but aggregation takes much longer due to the increase in the distinct attribute values.
 For difference ($\Delta$), we extend $\mathcal{T}_{old}$ with union semantics, while $\mathcal{T}_{new}$ is our reference point, and we present our results at logarithmic scale.
 In Fig. \ref{fig:operdblp.3}, we consider $\mathcal{T}_{old}$ - $\mathcal{T}_{new}$. As $\mathcal{T}_{old}$ expands, time increases as the output of the operation also grows. Regarding static attributes, similar to intersection, difference requires more than double the time required for both types of aggregation, whereas for the time-varying attribute aggregation is more expensive. Figure \ref{fig:operdblp.4} depicts $\mathcal{T}_{new}$ - $\mathcal{T}_{old}$, that needs less time compared to $\mathcal{T}_{old}$ - $\mathcal{T}_{new}$, as the operation output decreases. Any aggregation is faster than the operation for both the static and time-varying attribute, as it is actually time point aggregation.

Figure \ref{fig:opertri} depicts the aggregation time of the union ($\cup$) and intersection ($\cap$) for the static attribute in \textit{tri-DBLP}. Besides the times for the operation, and aggregation, we also present the time for the  the tri-graph construction ($G$-$tri$). For union (Fig. \ref{fig:opertri.1}), we first construct the tri-graphs and then apply union on them as this approach records a 2.5x speedup compared to the alternative that would need to compute the triangles in a very large union graph. The tri-graph construction is the most time-consuming requiring from 31s for the shortest interval up to 9k for the longest one. For intersection (Figure \ref{fig:operdblp.2}), we employ the second aggregation method, first creating the intersection graph and then computing and aggregating the tri-graph, as we register a 125x speedup in comparison to the first approach, since intersection reduces the size of the graph on which we need to compute the triangles. Here, aggregation  needs the most time up to 0.71s in [2000, 2002], while the operation is similar on all intervals with average time equal to 0.02s, and the creation of the triangles does not exceed 0.14s.

\noindent\textit{\textbf{Partial Materialization.}}
We evaluate our proposed optimizations by measuring speedup defined as the execution time of the proposed aggregation algorithms to the time of optimized aggregation that exploits precomputed results. In Fig. \ref{fig:spupu}, we report speedup when exploiting precomputed aggregations on time points to derive their non-distinct union compared to computing it from scratch. For \textit{DBLP}, aggregation over the years offers an 8x to 20x speedup for static attributes (Fig. \ref{fig:spupu.1}), and from 8x up to 78x for time-varying ones (Fig. \ref{fig:spupu.2}). For the gender attribute of \textit{DBLP-tri} in Fig. \ref{fig:spupu.3}, we record up to 252Kx speedup. The high speedups achieved in this case are due to the significant amount of time required for the creation of the triangle-based graph of \textit{DBLP}, while the optimized aggregation time is only slightly affected by the size of the graph.

Figure \ref{fig:spupdim} depicts speedup when exploiting precomputed attributes aggregations to derive aggregations on subsets of those attributes compared to computing them from scratch. Figure \ref{fig:spupdim.1} shows gender and publications aggregation when computed from the aggregation of all DBLP attributes with a speedup of 6x up to 21x. For \textit{MovieLens}, first we report each attribute when computed from all pairs of attributes (Fig. \ref{fig:spupdim.2}). In particular, gender is computed in $G1$ from (gender, age), in $G2$ from (gender, rating) and in $G3$ from (gender, occupation), and similarly for rating, with $R1$ from (rating, gender), $R2$ from (rating, age) and $R3$ from (rating, occupation), where we achieve a significant speedup of up to 48x. For \textit{MovieLens}, we also report results for computing all pairs (Fig. \ref{fig:spupdim.3}) and triplets (Fig. \ref{fig:spupdim.4}) of attributes from the aggregate of all 4 attributes, where we observe up to 8x and 6x speedup, respectively. Improvement is greater for single attributes followed by pairs and then, triplets. Finally, in Fig. \ref{fig:spupdim.5}, we present the speedup for the triangle-based graph of \textit{DBLP}, where similarly to efficient union, we record high speedups up to 210Kx.


\subsection{Qualitative Evaluation}
In this set of experiments, we focus on the Primary School dataset, and supposing we want to study disease spread dynamics, we perform a qualitative study of its evolution.  

Figure \ref{fig:f14.1} depicts the evolution graph on the aggregation for gender, omitting nodes of unspecified gender (U), from the 3\textsuperscript{rd} hour, lesson time, to the 4\textsuperscript{th} hour, break time. For each graph element, we provide the distribution for the events of stability, growth and shrinkage, and the total number of events. Similarly, Fig. \ref{fig:f14.2} presents the corresponding evolution graph for \textit{Primary School-tri}. It seems that evolution does not depend on gender. Comparing the two graphs, we see that edges between females (F-F) exhibit 16.6\% stability, which is about 5x higher compared to triangles of females (FFF) with 2.9\% stability. Similarly, for males there is a stability of 16.6\% for M-M edges and 3.1\% for MMM triangles, showing that relationships between larger groups of students have much smaller stability regardless of their gender, showing little opportunity for isolation bubbles.  

Next, we focus on F-F interactions. For stability with maximal interval pairs, we gradually decrease the maximum weight ($w_{th} = 242$), setting $k_3 = w{th}$, $k_2 = w{th}/2$ and $k_1 = w{th}/8$. We observe the highest stability for reference point 12 w.r.t. 11, denoted as (12, [11, 11]) where there are at least 242 common edges between girls. Figure \ref{fig:f15.2} depicts growth with minimal interval pairs, where $w_{th} = 342$ and $k_3 = w{th}$, $k_2 = w{th}/2$ and $k_1 = w{th}/4$. We observe the greatest growth on the 2\textsuperscript{nd} hour w.r.t. to the 1st with at least 342 new edges. For shrinkage and minimal intervals pairs with $w_{th} = 47$ which is gradually increased, setting $k_1 = w{th}$, $k_2 = w{th}*4$ and $k_3 = w{th}*8$, we observe the greatest shrinkage on (12, [8, 11]), where 47 edges are eliminated (Fig. \ref{fig:f15.3}). From the above, we can derive that the 2\textsuperscript{nd} time point appears to be of the highest risk for disease spread, while the 12\textsuperscript{th} reports high rates of stable and deleted interactions between girls.

\begin{figure*}
\centering
\begin{subfigure}{0.24\textwidth}
\includegraphics[width=\textwidth]{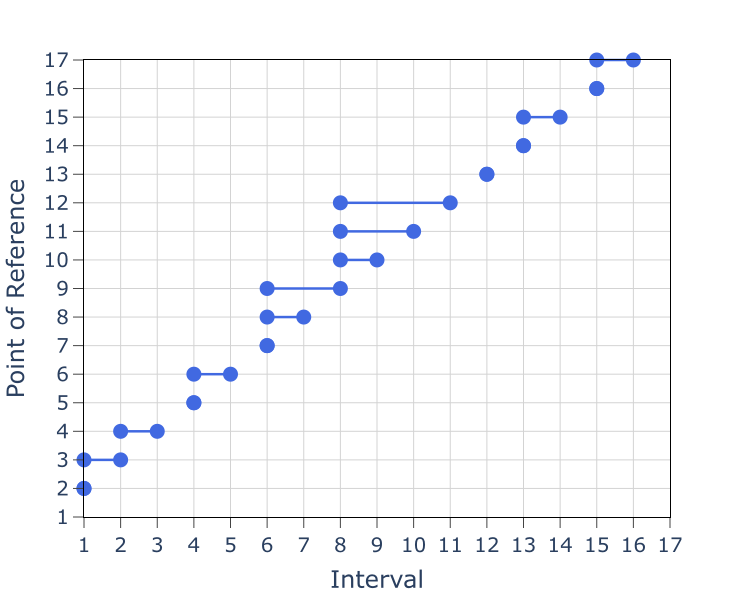}
\caption{Girls interactions}
\label{fig:pm12.1}
\end{subfigure}%
\begin{subfigure}{0.24\textwidth}
\includegraphics[width=\textwidth]{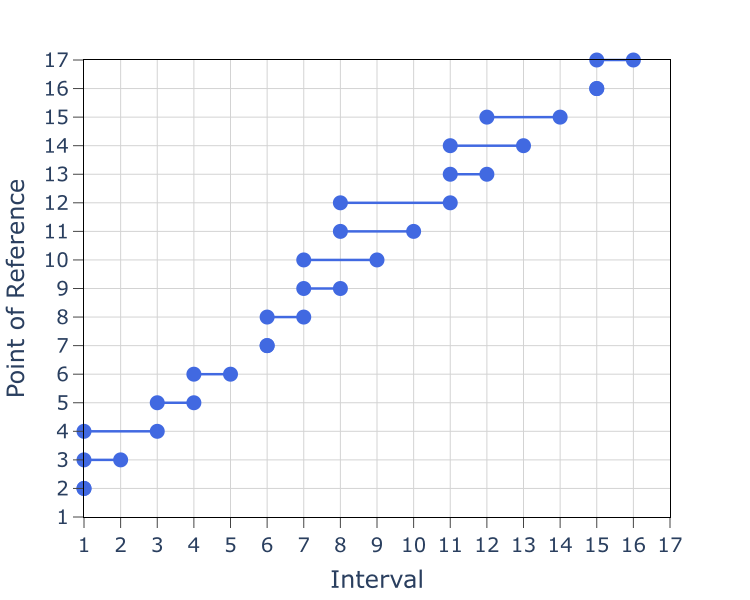}
\caption{Boys interactions}
\label{fig:pm12.2}
\end{subfigure}%
\begin{subfigure}{0.24\textwidth}
\includegraphics[width=\textwidth]{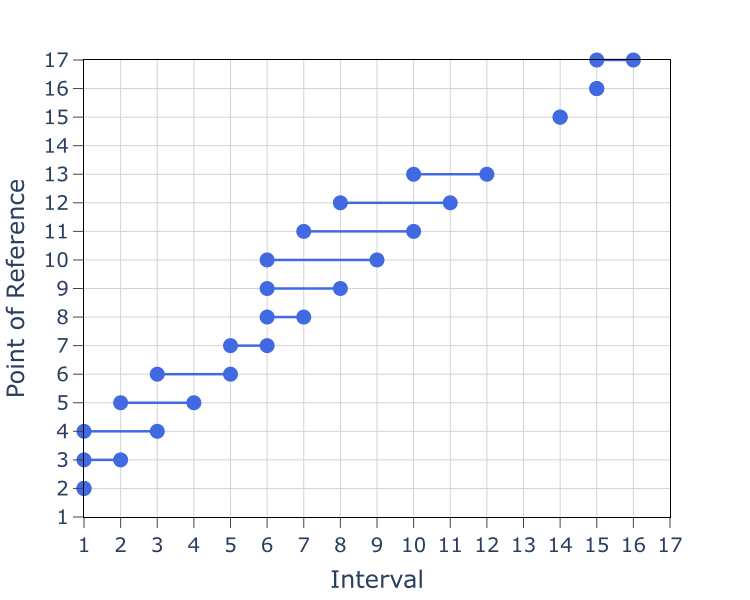}
\caption{1A's children interactions}
\label{fig:pm12.3}
\end{subfigure}%
\begin{subfigure}{0.24\textwidth}
\includegraphics[width=\textwidth]{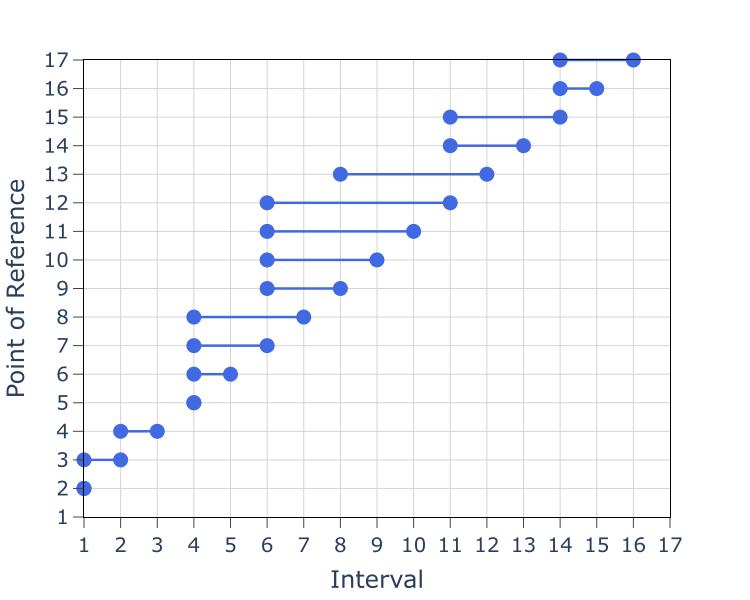}
\caption{5A's children interactions}
\label{fig:pm12.4}
\end{subfigure}%
\caption{Stability event for interactions between children with the same attribute value for \textit{Primary School}.}
\label{fig:pm12}
\end{figure*}

Finally, we further explore stability (Fig. \ref{fig:pm12}). First, we focus on gender and compare the exploration results for $k=50$ stable contacts between girls (fig. \ref{fig:pm12.1}) and boys (fig. \ref{fig:pm12.2}). The highest stability for girls is on (12, [8, 11]), (9, [6, 8]) and (11, [8, 10]), while for boys on (12, [8, 11]), (4, [1, 3]), (10, [7, 9]), (11, [8, 10]), (14, [11, 13]) and (15, [12, 14]), showing that boys interactions are more stable in comparison to girls. Next, we study the class attribute, and $k=10$ stable contacts. In Fig. \ref{fig:pm12.3} for students of junior class (1A), the highest stability is in (10, [6, 9]), (11, [7, 10]) and (12, [8, 11]). For students of a senior class (5A), we observe that a long interval (12, [6, 11]) indicates the highest stability, so we derive that older students have more stable contacts compared to younger ones (Fig. \ref{fig:pm12.4}). Consequently, boys seem to build more stable connections compared to girls and senior students appear to be of lower risk compared to junior ones for disease spread.

\section{Related Work}
With respect to related work, we discern between works that 
 focus on graph OLAP that ignores time, temporal relational data and temporal graph OLAP.

\noindent\textbf{\textit{Graph OLAP.}}
In \cite{Chen08, Chen09}, dimensions, measures and operators for graph OLAP are defined. Aggregate graphs are defined on both attribute and structural dimensions and roll-up, drill-down, slice and dice are supported. Partial materialization is discussed, while efficient computation for structural OLAP is also addressed in \cite{Qu11}. In \cite{Tian08}, drill-down and roll-up are supported on graphs aggregated based on node attributes and relationships between them.  GraphCube \cite{Zhao11} also defines aggregate graphs based on both attributes and structure, where the evaluated measure is the actual aggregated graph, while it extends OLAP with queries between aggregated graphs. Graphoids are defined in \cite{Gomez20} over labeled directed multi-hypergraphs to deal with heterogeneous data that are connected with more than binary relationships. In \cite{Diao21}, aggregates on RDF graphs are based on SPARQL aggregation semantics, and the problem of efficiently determining the most interesting attribute aggregations is addressed. 
In \cite{Deutch20}, TigerGraph's query language defines accumulators for aggregating values returned by pattern matching. State information is gradually computed for nodes accumulating the corresponding aggregate values for a query. In \cite{LiGuan13}, aggregation is based on random walks. Each node has a score based on the concentration of the attribute values of nodes in its vicinity. The goal is to determine interesting nodes based on a user-defined threshold. For more, a comprehensive review on graph OLAP  can be found in \cite{Queiroz-Sousa20}. Also, as graph aggregation provides a concise graph representation, it is also deployed for graph summarization \cite{Cebiric19}. 
All works above do not focus on temporal data, and while various types of graph aggregation including attribute aggregation were previously defined, they cannot support time-varying attributes or varying temporal resolution without significant changes to their methods and increase in their cost.

\noindent\textbf{\textit{Temporal Relational OLAP.}}
There is a lot of work in temporal relational databases.
Temporal relational algebra \cite{Dey96} employs multidimensional tuple timestamping and supports both valid and transactional time. Relational operators are extended to preserve snapshot semantics, and temporal projection and group by are defined. In \cite{Dohr18}, new temporal operators on sets of intervals are implemented to overcome the complexity of temporal calculations. A survey \cite{Golfarelli09} on time management in data warehouses covers several issues such as warehouse design, querying and schema changes.

\noindent\textbf{\textit{Temporal Graph OLAP.}}
Works on incorporating time in the graph OLAP model introduce new temporal models and operators. In \cite{Debrouvier21}, GQL is extended to T-GQL for handling temporal paths. Continuous paths address evolution and consecutive paths travel scheduling problems.  In \cite{Ghrab13}, a conceptual model with explicit labeling of graph elements is designed to support analytical operations over evolving graphs, and particularly time-varying attributes. This model incurs redundancy by adding elements that could be discovered through graph traversal, sacrificing performance for richer analysis support. In the  EvOLAP Graph \cite{Guminska18}, versioning is also used, both on attributes and graph structure to enable analytics on changing graphs but no temporal operators are supported. In contrast to our work, events are described as the difference of an entity state around a time point and do not capture evolution on periods of time. 
In TGraph \cite{Moffitt17}, temporal algebraic operators such as temporal selection for nodes and edges and traversal with temporal predicates are defined for a temporal property graph to capture the evolution of nodes, edges and their attributes on periods of validity. TGraph is further extended to allow aggregation on attributes and time  \cite{Aghasadeghi20}, while we also support pattern aggregation. Time aggregation focuses on viewing a graph in the appropriate resolution and studies stability, whereas, we focus on different events such as growth and shrinkage besides stability, and define an evolution graph that models all three through set based temporal operators. In \cite{Rost21, Rost22}, the authors present a tool that builds on the extensions of GRADOOP for temporal property graphs and supports graph summarization based on time and attribute aggregation. They provide different visualizations of a graph, i.e., the temporal graph view, the grouped graph view where grouped elements are enriched with both attribute and temporal aggregate information, and the difference graph view that illustrates new, stable and deleted elements between two graph snapshots. In contrast to our work, the system is driven by user queries and provides no exploration strategy to determine intervals with events of interest. In \cite{Andriamampianina22}, a conceptual model based on the property graph and the interval-based data model is utilized. This work focuses on the evolution of the graph both structurally and capturing capturing changes on the set of attributes and attributes’ values, though no temporal operators are used. While aggregation and evolution have been addressed, none of the above works provide any exploration strategies to detect important parts of a graph without depending on user-defined selections. 

Finally, the GraphTempo model for aggregating temporal attributed graphs is first introduced in \cite{Tsoukanara23}. Temporal and attribute aggregation is defined, along with the evolution graph and the two exploration strategies. In this work, we first introduce pattern aggregation by extending attribute aggregation to consider subgraphs that share common attribute values instead of individual nodes. We also define a new structure, the pattern graph, and present algorithms for using this structure to efficiently evaluate pattern aggregation as we experimentally show.

\section{Conclusions}
In this paper, we introduce GraphTempo, a model supporting temporal, attribute and pattern aggregation of evolving graphs. We define a set of temporal operators with both tight and relaxed semantics to support temporal aggregation and extend attribute aggregation to pattern aggregation that groups whole subgraphs instead of individual nodes. We also define a novel structure that captures graph evolution, and develop exploration strategies to determine minimal intervals with significant growth and shrinkage or maximal intervals with significant stability. We experimentally evaluate the efficiency of the proposed operators while also performing a qualitative evaluation based on three real datasets. We plan to extend the exploration strategies by considering top-$k$ and skyline queries to detect interesting events without requiring parameter configuration.

\section*{Acknowledgments}
Research work supported by the Hellenic Foundation for Research and Innovation (H.F.R.I.) under the “1st Call for H.F.R.I. Research Projects to Support Faculty Members \& Researchers and Procure High-Value Research Equipment” (Project Number: HFRI-FM17-1873, GraphTempo).

\bibliographystyle{abbrv}
\bibliography{bib}
\end{document}